\documentclass[12pt]{article}

\newcommand{\symp}{\omega}
\newcommand{\sect}{\sigma}
\newcommand{\dmsn}{n}
\usepackage[mathscr]{eucal}
\usepackage{amsmath,amsfonts,amssymb,amsthm}
\usepackage{times}
\usepackage{hyperref}

\advance\textheight\topskip
\numberwithin{equation}{section} \makeatletter
\@addtoreset{equation}{section}

\renewcommand{\hat}{\widehat}

\newcommand{\bref}[1]{\textbf{\ref{#1}}}

\renewcommand{\d}{\partial}
\renewcommand{\dh}{\mathrm{d_h}}
\newcommand{\dvv}{\mathrm{d_v}}
\newcommand{\derham}{\mathrm{d}}
\newcommand{\cM}{\mathcal{M}}
\newcommand{\cF}{\mathcal{F}}

\renewcommand{\geq}{\,{\geqslant}\,}
\renewcommand{\leq}{\,{\leqslant}\,}

\newcommand{\binner}[2]{%
  {\langle}\kern-4.15pt{\langle}#1{,}\,#2{\rangle}\kern-4.15pt{\rangle}}
\newcommand{\commut}[2]{[#1{,}\,#2]}

\newcommand{\pb}[2]{\left\{{}#1{},{}#2{}\right\}}

\newcommand{\ffrac}[2]{\raisebox{.5pt}%
  {\footnotesize$\displaystyle\frac{#1}{#2}$}\kern1pt}

\newcommand{\dl}[1]{\mathchoice{\ffrac{\d}{\d #1}}{\frac{\d}{\d #1}}{\ffrac{\d}{\d #1}}{\ffrac{\d}{\d #1}}}

\newcommand{\fR}{\mathbb{R}}

 \def\cH{\mathcal{H}}
 \def\cJ{\mathcal{J}}
 \def\cL{\mathcal{L}}

\usepackage{youngtab}
\usepackage{ytableau}
\usepackage[numbers,sort&compress]{natbib}
 \usepackage[nottoc]{tocbibind}

\makeatletter
\g@addto@macro\bfseries{\boldmath}
\makeatother

\usepackage{multirow}

\usepackage{bm}

\usepackage{hhline}

\usepackage{physics}

\usepackage{tikz-cd}

\usepackage{mathtools}

\usepackage{authblk}
\title{Presymplectic structures and intrinsic Lagrangians for massive fields\vspace{0.5em}}
\author[a,b]{Maxim Grigoriev\vspace{0.75em}}
\author[a,c]{Vyacheslav Gritzaenko\vspace{0.75em}}
\affil[a]{Lebedev Physical Institute,\protect\\
  Leninsky ave. 53, 119991 Moscow, Russia \vspace{1em}}
\affil[b]{Institute for Theoretical and Mathematical Physics,\protect\\
  Lomonosov Moscow State University, 119991 Moscow, Russia  \vspace{1em}}
\affil[c]{Moscow Institute of Physics and Technology,\protect\\
  Institutskiy per. 7, Dolgoprudny, 141700 Moscow region, Russia}
\date{}                     
\begin{document}

\maketitle

\begin{abstract}
We study the construction of the so-called intrinsic action for PDEs equipped with compatible presymplectic structures.  In particular, we explicitly demonstrate that the intrinsic action for the standard Einstein-Hilbert gravity is the familiar first-order Palatini action. Our main focus in this work is the massive spin-2 field, where the natural presymplectic structure is not complete in the sense that the associated intrinsic action does not reproduce all the equations of motion. We explicitly relate this feature to the differential consequences of the zeroth order in the genuine Lagrangian formulation of Fierz and Pauli. Moreover, a minimal multisymplectic extension of the intrinsic action that produces all the equations of motion is constructed and it is argued that systems of this type can be naturally regarded as multidimensional analogs of mechanical systems with constraints. Finally, we extend the considerations to the massive spin-3 field and argue that the extension to all the massive higher spins should be straightforward as well.

\end{abstract}


\newpage 
\tableofcontents

\section{Introduction}

In the context of modern (quantum) field theory one often encounters a problem of constructing a variational principle for a given system of partial differential equations (PDE). This is known as the inverse problem of the calculus of variations, see e.g.~\cite{Henneaux:1982iw,Henneaux:1984ke,Anderson1988} and references therein.  In its general form it implies constructing an explicit realization of a given PDE in terms of one or another set of dependent variables (=fields) together with the Lagrangian such that its  stationary surface (i.e. the equation manifold determined by the Euler-Lagrange (EL) equations and all their differential consequences) coincides with the initial PDE. 

PDEs can be defined in an invariant way without explicit reference to one or another set of dependent variables by specifying the equation manifold equipped with the involutive Cartan distribution~\cite{Vinogradov:1977} (for a review see e.g.~\cite{,Bocharov:1999,Krasil'shchik:2010ij}). In contrast to this, Lagrangian is a local horizontal top-form on the jet-bundle associated to the space of fields and hence can not be defined in terms of the intrinsic geometry of the equation manifold.  This implies that even to study possible Lagrangian formulations for a given PDE one is forced to consider explicit realizations of the PDE in terms of one or another set of fields. 

The very well-known example illustrating the intricacies of the inverse problem of the calculus of variations is the theory of massive spin-2 field in Minkowski space. Despite that the full set of the equations of motion for this system can not arise as Euler-Lagrange  equations (simply because the number of fields does not match the number of equations) the system is nevertheless Lagrangian. In their celebrated work~\cite{Fierz:1939ix} Fierz and Pauli demonstrated that by adding an additional field the system can be made Lagrangian. Moreover, the additional field is automatically set to zero due to the zeroth order  differential consequence of the EL equations while the equations on the remaining variables coincide with the initial PDE.

An invariant approach to the inverse problem is to work in terms of the geometric structures defined on the equation manifold (=stationary surface) that are capable of encoding the Lagrangian formulation. A natural candidate is the presymplectic structure (known also as symplectic current) induced by the Lagrangian on its stationary surface~\cite{Kijowski:1979dj,Crnkovic:1986ex,Zuckerman:1989cx,Anderson1991}.
Although in the case of ordinary differential equations (ODE) this approach turns out to be successful~\cite{Henneaux:1982iw} it fails in general in the case of PDEs. One of the issues being that in order to construct a Lagrangian out of a presymplectic structure one nevertheless needs to realize the equation as a surface in one or another jet-bundle and hence again face the ambiguity. Once PDE is realised in terms of  a given jet-bundle the presymplectic structure can be lifted to a Lagrangian~\cite{Khavkine2012,Sharapov:2016qne,Druzhkov:2021} whose EL equations are consequences of the initial PDE.

An alternative approach,  proposed in~\cite{Grigoriev:2016wmk},  is based on a distinguished realization of a given PDE as a surface in the jet-bundle of the equation manifold itself. Indeed, in terms of the intrinsic geometry of the PDE its solutions can be identified as parallel sections of the equation manifold considered as a bundle over the base manifold (space-time) and the connection being the Cartan distribution~\cite{Vinogradov:1977,Bocharov:1999}.\footnote{Such a representation of PDE is also known in the context of field theory under the name of unfolded formulation~\cite{Vasiliev:1980as,Lopatin:1987hz,Vasiliev:2005zu}. See also~\cite{Barnich:2010sw,Grigoriev:2012xg,Grigoriev:2019ojp} for the relation of the unfolded formulation to BV-BRST formalism on jet bundles and the geometrical approach to PDEs} The covariant constancy condition is a first order PDE on the components of the section.  Given a compatible presymplectic structure it determines a well-defined local functional on the space of sections, known as the intrinsic action. Under rather mild and purely technical assumptions it can be shown that the PDE determined by the intrinsic action either coincide with or weaker than the initial PDE. Moreover, it is possible to identify (though not in invariant terms) an easy to check criteria~\cite{Grigoriev:2016wmk} of whether a given Lagrangian system admits an equivalent 
formulation in terms of the intrinsic action. If it does the 
Lagrangian formulation is fully encoded in the compatible presymplectic structure on the equation manifold. Mention that the intrinsic action construction finds its roots in the presymplectic generalization~\cite{Alkalaev:2013hta,Grigoriev:2016wmk} of the celebrated AKSZ construction~\cite{Alexandrov:1995kv} for the Batalin-Vilkovisky formulation of topological models and, hence, it is not surprising that it has a far reaching generalization in the context of local gauge field theories~\cite{Grigoriev:2020xec}.

For most of the physically relevant examples of Lagrangian (gauge) field theories the intrinsic action defines an equivalent first-order Lagrangian formulation. In particular, in this work we explicitly demonstrate that in the case of Einstein gravity the intrinsic action determined by a natural symplectic structure induced by the Einstein-Hilbert action is precisely the familiar Palatini action. However,
it was observed already in~\cite{Grigoriev:2016wmk} that for massive spin-2 theory the intrinsic action is not complete in the sense that it doesn't reproduce all the equations of motion. The same phenomena also take place in the case of massive fields of spin higher than $2$.

In this work, after a brief review of the intrinsic action construction we concentrate on the example of massive spin-2 field. Our strategy is to explicitly demonstrate that the intrinsic Lagrangian is only partial, i.e. does not determine all the equations of motion and to identify the minimal extension of the intrinsic Lagrangian that has the same structure but
is equivalent to the Fierz-Pauli one.  This is achieved by employing the multidimensional generalization of the Ostrogradsky action (also known as parent action, see e.g.~\cite{Grigoriev:2010ic}) which provides a systematic way to equivalently rewrite any Lagrangian system in the multisymplectic form (also known as the covariant Hamiltonian form). By equivalently reducing this formulation via the elimination of the auxiliary fields but respecting its multisymplectic structure one arrives at the minimal extension of the intrinsic action. The structure of this extension suggests that in addition to the presymplectic structure the equation manifold carries an additional geometric structure such that together with the presymplectic one they determine a Lagrangian in a natural way. We also discuss possible interpretation of such systems as certain analogs of constrained Hamiltonian systems.

\section{Preliminaries}

The standard mathematical framework to analyse classical local field theories and their Lagrangian formulations is that of jet-bundles and their variations bicomplexes, see e.g.~\cite{Anderson1991}. Here we briefly recall the basic structures and statements that we need in this work.

Let $\cF\to X$ be a locally-trivial fiber bundle with base $X$ of dimension of $\dmsn$ and fiber $F$. Base space is interpreted as a space-time manifold and $F$ as a (locally defined) target space where the fields take values. A section $\sect:X \to \cF$ is interpreted as a field configuration. Introducing coordinates on the base $x^\mu$ and coordinates on the fibers $\phi^i$ section $\sect$ is locally determined by functions $\sect^i(x)=\sect^*(\phi^i)$. These are to be identified as component fields.

In order to study equations of motion for $\phi^i(x)$ and the associated Lagrangians it is extremely convenient to introduce jet-bundle $\mathcal{J} \equiv J^\infty(\mathcal{F})$
associated with $\cF$. This can be defined as a projective limit of finite jet-bundles $J^k(\cF)$. The point of the total space of $J^k(\cF)$ is a pair $(x,[\sect]_k)$, where $x\in X$ and 
$[\sect]_k$ is an equivalence class of sections of $\cF$ such that all their derivatives of order $l\leq k$ coincide at $x$ (although the equivalence relation explicitly employs coordinates it is coordinate independent). It follows $J^\infty(\cF)$ can be coordinatized by $x^\mu$, $\phi^i$, $\phi^i_\mu$, $\phi^i_{\mu \nu}$, $\ldots$, where e.g. $\phi^i_{\mu \nu}$ corresponds to $\d_\mu\d_\nu \sect^i(x)$.

Jet-bundle is equipped with the canonical Cartan distribution which assigns a horizontal completion to the vertical subspace at each point of the total space. In coordinate terms the distribution is determined by the following vector fields:
\begin{equation}
D_a = \frac{\d }{\d x^a} + \phi_a\frac{\d }{\d \phi} + \phi_{ab}\frac{\d }{\d \phi_a} + ...
\end{equation}
known as total derivatives. In this form it is clear that the distribution is involutive because $\commut{D_a}{D_b}=0$. One can also view the Cartan distribution as a flat Ehresmann connection on $\cJ$.

The decomposition of the tangent space into the direct sum of the vertical and horizontal subspaces induces an additional degree (horizontal form degree) on the algebra  $\bigwedge(\cJ)$ of local differential forms on $\cJ$ so that it decomposes as
\begin{equation}
    \bigwedge (\cJ)=\bigoplus_{0\leq k \leq \dmsn} \bigoplus_{l\geq 0} \bigwedge \nolimits^{(k,l)}  (\cJ)
\end{equation}
Elements of $\bigwedge^{(k,l)}(\cJ)$ are differential forms of horizontal degree $k$ and vertical degree $l$ or simply $(k,l)$-forms.

As basis horizontal differential forms one can take $dx^a$
and as vertical $\dvv \psi^i_{a_1\ldots}$, where by some abuse of notations $dx^a$ denotes $dx^a$ on $X$ pulled back to the total space by the bundle projection. 
In particular a generic $(k,l)$ form can be written as
\begin{equation}
 u = u_{a_1 \ldots a_k; {C_1 \ldots C_l} } (x, \psi)dx ^{a_1} ... dx ^{a_k} \dvv \psi ^{C_1} ... \dvv \psi^{C_l} \,,
\end{equation}
where  $\dvv \psi_{C}$ denotes  $\dvv \psi^i_{a_1\ldots}$, i.e. $C$ is a multi-index.

The decomposition of forms induces the decomposition of the de Rham differential
\begin{equation}
\derham = \dh + \dvv\,, \quad \dh:\bigwedge \nolimits^{(k,l)}(\cJ)\to \bigwedge \nolimits^{(k+1,l)}(\cJ)\,, \quad  \dvv:\bigwedge \nolimits^{(k,l)}(\cJ)\to \bigwedge \nolimits^{(k,l+1)}(\cJ)\,.
\end{equation}
In local coordinates $\dh$ is given by 
\begin{equation}
\dh = dx^a D _a\,.
\end{equation}
The algebraic relations between $\dh$ and $\dvv$ read as
\begin{equation}
\dh ^2 = \dvv ^2 = 0\,, \qquad  \dh \dvv + \dvv \dh = 0\,.
\end{equation}
The bigrading of $\bigwedge(\mathcal{J})$ makes it into the bicomplex, known as the variational bicomplex.


By definition, a system of partial differential equations
is (locally) given by a set of local functions $E_i \in \bigwedge^{(0,0)}(\mathcal{J})$ such that the surface it  defines is a subbundle of $\cJ$. An infinitely prolonged equation is the subbundle $\cM$ of $\cJ$ determined by 
\begin{equation}
    D_{a_1}\ldots D_{a_l}E^i=0\,, \qquad l=0,1,\ldots\,.
\end{equation}
Because by construction Cartan distribution on $\cJ$ is tangent to $\cM$ it defines an involutive distribution on $\cM$ and makes $\bigwedge(\cM)$ into a variational bicomplex. 

It is known that $\cM$, seen as a fiber bundle over $X$ 
equipped with the Cartan distribution, defines the equation understood as an invariant geometrical object. Because Cartan distribution determines horizontal differential $\dh$ on $\bigwedge(\cM)$ and  other way around, we denote PDE by $(\cM,\dh)$. In particular, solutions of $(\cM,\dh)$ are sections $\sect:X\to \cM$ to which the Cartan distribution is tangent (or, equivalently, covariantly constant sections) see e.g.~\cite{Krasil?shchik-Lychagin-Vinogradov,Krasil'shchik:2010ij}. If $\psi^A$ are local coordinates on the fibres of $\cM$ a section is parameterized by the functions $\sigma^A(x)=\sigma^*(\psi^A)$ and hence seen as a submanifold of $\cM$ the section is locally singled out by the constraints $\psi^A- \sigma^A(x)$. The condition that $D_a$ is tangent to the section then reads
\begin{equation}
\label{intr-PDE}
\d_a \sigma^A(x)-\Gamma_a^A(\sigma(x),x)=0\,,\qquad D_a=\d_a+\Gamma_a^A(\psi,x)\dl{\psi^A}\,,
\end{equation}
where the second formula determines  "connection coefficients" $\Gamma_a^A$ in terms of the total derivatives seen as locally defined vector fields on $\cM$. In more invariant terms the condition that $\sigma$ is a solution reads:
\begin{equation}   
\label{theory_cons}
    d\circ \sigma^*=  \sigma^* \circ \dh\,.
\end{equation}
Applying both sides to $\psi^A$ one indeed recovers~\eqref{intr-PDE}. It is important to stress that the above form gives an equivalent representation of the initial PDE as a first-order PDE and this representation is defined solely in terms of intrinsic geometry of the equation manifold $\cM$. We refer to this as to the intrinsic representation. In the context of field theory formulations of this type are often called unfolded, see~\cite{Vasiliev:2005zu} and references therein.

System of PDE $\{E_i\}$ defined on $\cJ$ is called Euler-Lagrange (EL) if there exists a local $(\dmsn, 0)$ form $\mathcal{L}$ such that 
\begin{equation}
E_i = \frac{\delta^{EL} \cL}{\delta\phi^i}\,, \qquad \frac{\delta^{EL} \cL}{\delta\phi^i} \equiv \frac{\d \mathcal{L}}{\d \phi ^i} - D _a \frac{\d \mathcal{L}}{\d \phi ^i _a} + D _a D _b \frac{\d \mathcal{L}}{\d \phi ^i _{ab}} - ...
\end{equation}
The operation defined by the second equality is known as Euler-Lagrange derivative. It can be defined more invariantly as the operation satisfying:
\begin{equation}
\dvv \cL=\dvv \phi^i \frac{\delta^{EL} \cL}{\delta\phi^i} - \dh \hat{\chi}\,, \qquad 
\frac{\delta^{EL} (\dh \rho) }{\delta\phi^i}=0\,\,\, \forall \rho\,.
\end{equation}
It is also convenient to introduce Euler differential $\delta^{E}=\dvv \phi^i \frac{\delta^{EL} }{\delta\phi^i}$.

It is clear that the property of a PDE, whether it be Euler-Lagrange or not, is not invariant under the equivalence. For instance, if $E_i$ are Euler-Lagrange equations (i.e. $E_i=\frac{\delta^{EL} \cL}{\delta\phi^i}$), the equivalent equations $E^\prime_i=\Lambda_i^j E_j$, where $\Lambda_i^j$ is an invertible local operator, are not Euler-Lagrange in general. Identification of equations equivalent (in this sense) to Euler-Lagrange ones is a well-known multiplier problem \cite{Khavkine2012, Henneaux:1984ke}. More generally, two PDEs are called equivalent if the respective equation manifolds are isomorphic as bundles over $X$ and the isomorphism identifies the respective Cartan distributions. It is natural to call PDE $(\cM,\dh)$ Lagrangian if there exists a jet-bundle $\cJ$ and a local form $\cL\in\bigwedge^{(\dmsn,0)}(\cJ)$, such that $(\cM,\dh)$ is equivalent to the infinitely prolonged equation determined by $E_i=\frac{\delta^{EL} \cL}{\delta\phi^i}$.

Given a Lagrangian $\cL\in\bigwedge^{(\dmsn,0)}(\cJ)$ it defines a presymplectic potential - an $(\dmsn-1, 1)$-form on $\mathcal{J}$ defined through:
\begin{equation}
\label{theory2}
\dvv \mathcal{L} = \dvv \phi ^i E_i - \dh \hat{\chi}
\end{equation}
The ambiguity in $\hat \chi$ is given by $\dh$-closed forms and hence (locally) $\dh$-exact ones. Presymplectic potential $\hat \chi$ determines the presymplectic form $\hat{\symp} = \dvv \hat{\chi}$:
\begin{equation}
\hat{\symp} = \dvv \hat{\chi} = d \hat{\chi} - \dh \hat{\chi} = \derham (\hat{\chi} + \mathcal{L}) - \dvv \phi E\,.
\end{equation}
If $\symp,\chi$ denote $\hat\symp,\hat\chi$ pulled-back to $\cM$ one finds:
\begin{equation}
\label{symp_1}
\symp = \derham (\chi + \mathcal{L}|_\cM)\,.
\end{equation}

Consider as an example a system whose Lagrangian does not involve derivatives of order higher than 2. Then $\hat{\chi}$ is given explicitly by:
\begin{eqnarray}
\label{theory}
\hat\chi =  ((\frac{\d \mathcal{L}}{\d \phi_a} - D _b \frac{\d \mathcal{L}}{\d \phi_{ab}})\dvv \phi + \frac{\d \mathcal{L}}{\d \phi_{ab}} \dvv \phi_b) (dx)^{d-1} _a\,.
\end{eqnarray}
Here and in what follows we use:
\begin{equation}
    (dx)^{\dmsn-k}_{a_1\ldots a_k}= \frac{1}{(\dmsn - k)!} \epsilon _{a_1 \ldots a_k c_1 \ldots c_{\dmsn - k}} dx ^{c_1} \ldots dx ^{c_{\dmsn - k}}
\end{equation}

\section{Intrinsic action}
\subsection{Multisymplectic systems}\label{sec:multisymp}

Consider a fiber bundle $F\times X \to X$ which for simplicity we assume trivial and finite-dimensional.
The algebra of local forms on $F\times X$ decomposes with respect to the vertical and the horizontal form degree and hence is a bicomplex so that de Rham differential can be represented as $d=d_X+d_F$ (where $d_X$ can be also regarded as horizontal while $d_F$ as the vertical differential). If $x^a$ and $\psi^A$ be local coordinates respectively $X$ and $F$ they give a natural coordinate system on $F\times X$ the decomposition of the de Rham differential reads as: $d=d_X+d_F=dx^a\dl{x^a}+d\psi^A\dl{\psi^A}$.

Suppose that $F\times X$ is equipped with $(\dmsn-1, 1)$ form $\overline{\chi}$ and $(\dmsn,0)$ form $\overline{\mathcal{H}}$. This data defines a natural action functional on the space of sections. More precisely, if $\sigma:X \to F\times X$ is a section 
\begin{equation}
\label{multisymp}
S[\sigma] = \int \sigma^*(\bar\chi)-\sigma^*(\bar{\cH})\,.
\end{equation}
In terms of coordinates, $\sigma$ is determined by fields $\psi^A(x)=\sigma^* (\psi^A)$ and the explicit form of the action reads as:
\begin{equation}
\label{multsymp2}
S[\psi]=\int (d \psi ^B \bar\chi_B(\psi(x),x,dx))-\bar\cH(\psi(x),x,dx)\,.
\end{equation}
The Lagrangian system determined by the above data is often called multisymplectic, see e.g.~\cite{Gotay:1997eg,Hydon:2005,Bridges2009, Ibort:2016}. One may also notice that (extended) Hamitonian action of the (constrained) Hamiltonian system is also of the form \ref{multisymp} so that \ref{multisymp} can be considered a multidimensional generalization of the Hamiltonian action with constraints and is often referred to as covariant Hamiltonian formulation. To simplify formulas, in what follows we make the following technical assumption: $\chi$ satisfies $\dh \chi=dx^a\d_a \chi=0$.

As we review in Section~\bref{sec:parent} any Lagrangian system can be systematically represented in the multisymplectic form at the price of introducing auxiliary fields. Moreover, for most of the usual examples of (gauge) theories there exists a multisymplectic formulation such that:
\begin{enumerate}
    \item Among the EL equations of~\eqref{multisymp} there are no algebraic relations on $\psi^A$. More precisely,  variables $\psi^A$ (seen as coordinates on the respective jet-bundle) remain independent when restricted to the stationary surface of~\eqref{multisymp}.  
    
    \item \label{extra} The $(n-1,2)$-form $\bar\omega=d_F\chi$ is nondegenerate everywhere in the following sense: if $V$ is a vertical vector at a given point and $i_V\omega=0$ then $V=0$. 
\end{enumerate}
The 2-form $\bar \omega$ can be written in local coordinates as $d \psi^A d \psi^B\bar \omega^c_{AB}(\psi,x)(dx)^{n-1}_c$ and the nondegeneracy property reads as: $\omega^c_{AB}V^B=0$ implies $V^B=0$. Following~\cite{Grigoriev:2016wmk} we call a Lagrangian system natural if it can be equivalently reformulated in the multisymplectic form satisfying the above two additional conditions. Note however, that in~\cite{Grigoriev:2016wmk} the second condition was reformulated in an equivalent way in terms of algebraic gauge symmetries.

The equations of motion of a multisymplectic system read
as:
\begin{equation}
\bar\omega_{AB}(\psi(x),x,dx)d\psi^B(x)-(\d_A \bar H)(\psi(x),x,dx)=0\,.
\end{equation}
If $\bar\omega$ is nondegenerate these can be rewritten as:
\begin{equation}\label{mult-eom-2}
\bar \omega_{AB}(\psi(x),x,dx)\left(d\psi^B(x)-Q^B(\psi(x),x,dx)\right)=0\,,
\end{equation}
for some $Q^B(\psi,x,dx)=dx^a Q_a^B(\psi,x)$, which are not unique, in general.

The nondegeneracy of $\bar\omega$ implies that the Lagrangian of the multisymplectic system does not have algebraic gauge symmetries (also known as Stueckelberg symmetries). In the case at hand an algebraic gauge transformation can be defined as that of the form
\begin{equation}
\delta \psi^A(x)=R^A(\psi,x)\epsilon(x)\,,
\end{equation}
with $R^A$ such that $R^A(\psi,x)\epsilon(x)=0$ implies $\epsilon(x)=0$. In fact, an even stronger statement holds: the following conditions are equivalent: (i)  $\bar\omega$ is nondegenerate; (ii) Lagrangian~\eqref{multisymp} does not have algebraic gauge symmetries. This is easily seen using~\eqref{mult-eom-2}. In particular, this shows the equivalence of the  above definition of the natural system and that from~\cite{Grigoriev:2016wmk}.

\subsection{Intrinsic action}\label{sect:intrinsic}

A natural multisymplectic Lagrangian can be associated to an equation manifold equipped with a compatible presymplectic structure. More precisely, suppose we are given with a PDE $(\cM,\dh)$ such that the respective equation manifold $\cM$, seen as a bundle over the space-time $X$, is equipped with a compatible presymplectic form $\symp$, that is $(\dmsn-1, 2)$-form satisfying $\dh\symp=\dvv\symp=0$. 
It follows there exist $\chi,l$ such that $\symp = \derham (\chi + l)$, where $l$ is an $(\dmsn, 0)$ form and $\chi$ is a presymplectic potential $(\dmsn-1,1)$-form, $\dvv \chi = \symp$. More precisely, $l$ can be found from $\dh \chi = - \dvv l$ because $\dvv$ is locally acyclic.  For a Lagrangian system with the Lagrangian $(\dmsn,0)$-form $\mathcal{L}$ form $l$ can be taken as $\mathcal{L}|_\cM$.  Note, however, that we do not require that $(\cM,\dh)$ is necessarily Euler-Lagrange and that $\symp$ necessarily arises from a Lagrangian.

Now, following~\cite{Grigoriev:2016wmk}, consider a new field theory whose fields are sections of $\cM$. If $\sect:X\to \cM$ is a section then one defines the following action functional:\footnote{Note that if we disregard the decomposition of $\chi+l$ with respect to horizontal and vertical form degree $\chi+l$ is just a form of total form degree $n$ and the action has a clear geometrical meaning. Note also that if we extend our bundle $\cM$ to a bundle over an $n+1$ dimensional manifold $\bar\cM$ whose boundary is $\cM$, the action can be rewritten in the WZW-like form $\int_{\bar\cM}\sigma^*(\omega)$. This coincides with~\eqref{intr-action} via Stocks formula and $d\omega=\chi+l$.}
\begin{equation}
\label{intr-action}
  S^C[\sect] = \int \sect^*(\chi+l)\,.
\end{equation}
Using adapted coordinate system $\psi^A,x^a$ on $\cM$, where $x^a$ are coordinates on the base pulled back to $\cM$, introduce a $(\dmsn,0)$-form $\cH = \dh \psi ^A \chi_A - l$. The above action takes the form:
\begin{equation}
\label{intr-action2}
  S^C[\sect] = \int \sect^* (\derham\psi^A \chi_A) - \sect^*(\mathcal{H}) = \int\derham(\sect^* (\psi^A)) \sect^*(\chi_A) - \sect^*(\mathcal{H})\,.
\end{equation}
In components it reads as:
\begin{equation}
\label{theory5}
  S^C[\psi] = \int (d\psi^A(x) \chi_A(\psi(x),x,dx)  - \mathcal{H}(\psi(x),x,dx))\,,
\end{equation}
where by some abuse of notations $\psi^A(x)=\sigma^*(\psi^A)$.

The variation of the action under the infinitesimal variation $\delta \sigma$
is given by
\begin{equation}
\label{variation}
    \delta S^C=\int \delta \psi^A \omega_{AB}(\psi(x),x,dx)(d\psi^B(x)-Q^B(x,dx))+\text{boundary terms}\,,
\end{equation}
where $Q^B(\psi(x),x,dx)=\sigma^*(\dh\psi^B)$. In particular, the EL equations read explictly as
\begin{equation}
    \omega_{AB}(\psi(x),x,dx)(d\psi^B(x)-Q^B(x,dx))=0
\end{equation}
and are the consequences of the intrinsic form~\eqref{intr-PDE} of the  PDE under consideration.  The action functional defined on sections of $\cM$ by~\eqref{intr-action} is referred to in what follows as the intrinsic action. Its advantage is that it is defined in terms of the intrinsic geometry of the equation manifold $\cM$. More precisely, it is determined by the  presymplectic form $\symp$ and the horizontal differential $\dh$ defined on $\cM$.

The crucial point is the interpretation of the intrinsic action. Despite the fact that $\cM$ is generically infinite-dimensional the intrinsic action depends only on a finite number of coordinates because $\chi$ is local. In order to give an intrinsic action an unambiguous interpretation it is natural to gauge fix those fields on which the action does not depend.  More precisely, suppose that one has found a set of linearly independent vertical vector fields $R_\mu$ on $\cM$ such that $\symp _{AB}R_\mu^A = 0$ and $R_\mu$ form a basis in the vertical kernel distribution of $\symp$. These vectors define gauge transformations preserving the above action. Indeed, setting $\delta\psi^A=R^A_\mu\epsilon^\mu$, where $\epsilon^\mu(x)$ are arbitrary gauge parameters, \eqref{variation} implies that $\delta_\epsilon S^C=0$ vanishes modulo boundary terms.

The distribution on $\cM$ determined by vector fields $R_\mu$ is by construction involutive. Indeed, if $R_\mu$ is a maximal set of linearly independent vertical vectors $R_\mu$ on $\cM$ such that $i_{R_\mu} \symp = 0$ it follows that $i_{[R_\mu, R_\nu]} \symp = 0$. Here we regard $\omega$ as a vertical form with values in horizontal $n-1$-forms. Thereby, assuming regularity we can at least locally find new vertical coordinates $\phi^\alpha, \psi^i$ such that $R_\mu = R_\mu ^\alpha \frac{\d}{\d \phi ^\alpha}$ with $R_\mu^\alpha$ invertible. It follows the gauge transformation for  $\phi^\alpha$ can be equivalently represented as $\delta \phi^\alpha = \epsilon^\alpha$,
where  $\epsilon^\alpha=\epsilon^\alpha (x)$ are arbitrary functions and hence $\phi^\alpha$ can be gauge-fixed by e.g. setting  $\phi^\alpha =0$, giving the Lagrangian system with fields $\psi^i(x)$ and the action given by ~\eqref{intr-action} with $\phi^\alpha=0$.\footnote{An alternative and probably more fundamental interpretation of the action can be achieved by resorting to the graded geometry and BV-BRST formulaion. More precisely, extending $\cM$ to a bundle $E$ over $T[1]X$ the local functions on $E$ can be identified as horizontal local forms on $\cM$ while $\omega$ becomes a presymplectic 2-form of degree $n-1$. Now consider the space of supersections $T[1]X \to E$. Presymplectic form $\omega$ naturally defines  a vertical presymplectic structure $\Omega^E$ of degree $-1$, see~\cite{Alexandrov:1995kv,Grigoriev:2020xec}. Taking a symplectic quotient results in a symplectic structure which by construction involves only coordinates of degree $1,-1$ and can be interpreted as a Batalin-Vilkovisky (BV)  symplectic structure associated to the intrinsic action. The intrinsic action is well-defined on the symplectic quotient. At the technical level this procedure is a minor variation of that explained in~\cite{Grigoriev:2020xec} and it gives a BV description of the system. Note however, that we have not incorporated the information about gauge invariance and hence the BV action coincides with the classical one, i.e. does not depend on antifields.}

The equations of motion of the new Lagrangian system can be either equivalent or not equivalent to the initial PDE $(\cM,\dh)$. In the former case we call a presymplectic structure $\symp$ complete while in the later partial or weak. Roughly speaking, for a complete $\symp$ the equations  in \eqref{theory_cons} that are complementary to the equations for $\psi^i$  express $\phi^\alpha$ in terms of $\psi^i$ and their derivatives.

Natural Lagrangian systems give rise to complete presymplectic structures~\cite{Grigoriev:2016wmk}. Indeed, starting with the multisymplectic formulation satisfying the extra two conditions stated in section~\bref{sec:multisymp} one finds that the presymplecic structure determined by the Lagrangian gives back the initial multisymplectic action via the intrinsic Lagrangian construction (see Section~\bref{sec:mult-ex} for more details). More precisely, 
among the equivalence class of presymplectic structures determined by the Lagrangian, one picks one that depend on 0-th jets (i.e. involves only undifferentiated fields). In other words, for natural Lagrangian systems the Lagrangian formulation is entirely encoded in the intrinsic geometry of the equation itself. 
However, not all interesting Lagrangian systems are natural. In this work we study such systems and show that they can be regarded as systems with constraints.

\section{Intrinsic Lagrangians of natural systems: examples}
Before considering Lagrangian systems that are not natural let us first illustrate the intrinsic Lagrangian construction on the examples of natural systems, arising in mechanics and field theory.

\subsection{ODE system}
Let us discuss the simplest example, mechanics. Let $F$ be a finite-dimensional phase space of a mechanical system and let $z^a$ be the local coordinates. The equations of motion can be represented as
\begin{equation}
\label{em_mech}
\dot{z}^a = V^a(z(t),t)
\end{equation}
for some $V^a (z, t)$. Geometrically, $V^a$ are components of a vertical vector field on the trivial fiber bundle $F\times \fR^1$.

The equation manifold can be identified with  $\cM=F\times \fR^1$ itself and is a trivial bundle over $\fR^1$. The horizontal differential (on the equation manifold) for this system is explicitly given by:
\begin{equation}
\dh = dt(\frac{\d}{\d t} + V^a \frac{\d}{\d z^a})\,.
\end{equation}

Suppose that $\cM$ is equipped with a nondegenerate vertical presymplectic form $\symp = \symp _{ab}(z,t) \dvv z^a \dvv z^b$ (note that because the space-time dimension $\dmsn=1$ the form is purely vertical). Moreover, suppose that the
presymplectic form is compatible, i.e. $\dh\omega=0$. In the case where $\omega$ is $t$-independent (or, more geometrically, is a pullback from $F$) this amounts to $L_V\omega=0$, i.e. that the motion is canonical. 

In this case the intrinsic action is nothing but the usual Hamiltonian action given by 
\begin{equation}
S^C = \int dt (\dot{z}^a \chi_a  - H) \,,
\end{equation}
where $\chi$ is defined through $\symp = \dvv \chi$ and $H$ through $H=i_V\chi-l$ with $\dvv l+\dh \chi=0$. In the case where $\dl{t}\chi_a=0$ one gets $\dvv H =i_V \omega$, so that $H$ is indeed a Hamiltonian for $V$. Because $\symp_{ab}$ is invertible the EL equations of this action are precisely \eqref{em_mech}.   These considerations were originally put forward in \cite{Henneaux:1982iw}, where it was shown  that the existence of  a nondegenerate vertical presymplectic form is sufficient for the existence of the variational principle.

\subsection{Constrained mechanics}

Let $F$ be a phase space of the constrained Hamiltonian system. By definition, the dynamics of the system is governed by the extended Hamiltonian action given by: 
\begin{equation}
\label{constr1}
S = \int dt (\dot{z}^b \chi_b  - H - \lambda_\alpha T^\alpha)
\end{equation}
where function $T_\alpha(z)$ are constraints, $\chi$ is a symplectic potential, i.e. the symplectic 2-form $\symp=d\chi$, and $H$ is a Hamiltonian. The Poisson bracket on $F$ determined by $\omega$ is denoted by $\pb{\cdot}{\cdot}$. 

For simplicity we assume that the constraints are irreducible and are of the first class, i.e. that both $\pb{T^\alpha}{T^\beta}$ and $\pb{T^\alpha}{H}$ vanish on the constrained surface $\Sigma$ determined by $T^\alpha=0$. Moreover, we assume that the constraints are defined globally (or restrict the analysis to a suitable neighbourhood).

If among $T^\alpha$ there was a subset of second class constraints $T^{\alpha^\prime}$ we could have eliminates $T^{\alpha^\prime}$ together with their associated Lagrange multipliers $\lambda^\prime$ as auxiliary fields. This would results in the action of the same structure with $\chi, H$
being the initial $\chi,H$ pulled back to the surface $T^{\alpha^\prime}=0$ and new $\symp=\derham\chi$ invertible (because the surface $T^{\alpha^\prime}=0$ is second-class).

Now we apply the intrinsic Lagrangian construction to the above action. It is convenient to use a special coordinate system $y^\alpha,y^i$ on $F$, where $y^\alpha=T^\alpha$ and $y^i$ are complementary coordinates (their restriction to $\Sigma$ gives a coordinate system therein).
The equations of motion set $y^\alpha=0$ and
express $\dot y^i$ in terms of $y^i$ and $\lambda_\alpha$ in terms of $y^i,\dot y^i$. Because for a first class system there are no further differential consequences the stationary surface can be identified with $\cM=\Sigma\times \fR^1$ and as independent coordinates there one can take $t,y^i$ restricted to the surface. 

It is straightforward to check that the presymplectic potential induced by the Lagrangian \eqref{constr1} on its stationary surface is $\chi^\prime_i(y,t) dy^i$, where $\chi_i$ are coefficients of the 1-form $\chi$ pulled back to $\Sigma$, and the Hamiltonian is just $H^\prime=H|_\Sigma$. The intrinsic action is then given by:
\begin{equation}
\label{1st-red}
    S^{C}[y^i]=\int dt (\dot y^i\chi^\prime_i -H^\prime)\,.
\end{equation}
This is not a usual Hamiltonian action because in contrast to the initial $\symp=d\chi$ the 2-form  $\symp^\prime=d\chi^\prime$ is in general degenerate. More precisely, vector fields $R^\alpha=\pb{T^\alpha}{\cdot}|_{\Sigma}$ are in the kernel of $\symp^\prime$. These vector fields are just the generators of the gauge transformations determined by the first class constraints.  It is well known that under the usual regularity conditions (that $\symp$ is invertible and $T^\alpha$ are regular) these vector fields exhaust the kernel of $\symp^\prime$ on $\Sigma$, see e.g.~\cite{Henneaux:1992ig} for details. Because our stationary surface $M=\Sigma\times \fR^1$, these vector fields exhaust  the kernel of $\symp$ in the vertical subspace.

According to the interpretation of the intrinsic action we need to restrict to the gauge-fixing submanifold of $\Sigma$. If we disregard global geometry issues (as we do in this work) this is of course equivalent to passing to the symplectic quotient of $\Sigma$. If $u^i$ are coordinates on the quotient then the gauge-fixed intrinsic action takes the form:
\begin{equation}
\label{2nd-red}
    S^{C\prime}[u^i]=\int dt (\chi^{\prime\prime}_i \dot u^i-H^{\prime\prime})\,,
\end{equation}
where $\chi^{\prime\prime}$ and $H^{\prime\prime}$ are induced by $\chi^{\prime}$ and $H^{\prime}$ on the quotient (or equivalently are the initial $\chi,H$ pulled back to the gauge-fixing submanifold of $\Sigma$). Of course what we have arrived at is just the reduced phase space and the reduced phase space Hamiltonian action of the initial constrained system. 

The reduced phase space formulation~\eqref{2nd-red} can be obtained in various other ways, well-known in the literature. One possibility is to  immediately introduce gauge fixing conditions $G_\alpha$ such that $\pb{T^\alpha}{G_\beta}$ is invertible on $\Sigma$. It follows that the complete set $T_\alpha, G_\beta$ (understood as constraints) is second class and the above  reduced phase space action arises as that describing the reduced dynamics.

Another way is to eliminate $y^\alpha,\lambda^\beta$ as auxiliary fields and arrive at~\eqref{1st-red} by their elimination. One can then observe that the gauge transformation $\delta z^a=\pb{T^\alpha}{z^a}|_\Sigma \epsilon^\alpha$ determined by the 1st class constraints are
purely algebraic (Stueckelberg) for the coordinates along the kernel of $\symp^\prime$ and hence can be gauge-fixed algebraically resulting in~\eqref{2nd-red} in a suitable gauge.

A subtle point worth discussing here is that after eliminating $y^\alpha,\lambda_\beta$ as auxiliary fields, the gauge transformations induced by the first class constraints are purely algebraic. This may lead to a confusion because by purely algebraic operations (elimination of auxiliary fields and gauge-fixing algebraic gauge symmetries\footnote{These two operations are often unified under the name of elimination of generalized auxiliary fields~\cite{Dresse:1990dj} within BV formalism, where they have a unique homological interpretation, see e.g.~\cite{Grigoriev:2012xg} and refs. therein.}) the first class constrained system is equivalent to a non-gauge system (of course all this holds locally and under the usual regularity assumptions). This is a peculiarity of 1d systems. A local gauge field theory is in general not equivalent to the non-gauge theory via elimination of generalized auxiliary fields. For instance, in Maxwell theory there are nontrivial BRST cohomology classes~(see e.g.~\cite{Barnich:2000zw}) in positive ghost numbers, which are not related to global geometry, and hence are the obstructions to such an equivalence. At the same time for a mechanical first-class constrained system nontrivial BRST cohomology classes may only arise due to a global phase-space/constrained surface geometry or certain non-regularity of the constraints.

\subsection{A natural multisymplectic system}
\label{sec:mult-ex}

As we already discussed in Section~\bref{sect:intrinsic} natural Lagrangian systems admit complete presymplectic structures and hence their Lagrangian formulations are encoded in the presymplectic structure on the equation manifold. To illustrate this statement let us explicitly construct the intrinsic action for a natural system determined by the action~\eqref{multsymp2}. Using \eqref{theory} one finds a representative of the presymplectic potential:
\begin{equation}
\label{chimult}
\hat{\chi} = \bar\chi_A \dvv \psi^A
\end{equation}
Because the system is assumed natural, coordinates $\psi^A$ on its jet bundle remain independent when restricted to the equation manifold and hence together with $x^a$ can be completed to a coordinate system on the equation. In this coordinate system the explicit expression for the form $\hat\chi$ pulled back to the equation manifold remains unchanged, i.e. $\bar\chi _A \dvv \psi ^A$.

Furthermore, in this coordinate system the expression for the covariant Hamiltonian coincides with the initial $\bar H$. Indeed,
\begin{equation}
\cH = \bar\chi _A \dh \psi ^A - \cL (dx)^\dmsn = \bar\cH\,.
\end{equation}
It follows the intrinsic action is given by
\begin{equation}
S[\psi]=\int \left( d \psi ^B \bar\chi_B(\psi(x),x,dx)-\bar\cH(\psi(x),x,dx)\right)\,,
\end{equation}
where we gauged away all the remaining fields as they are in the kernel of the presymplectic structure $\dvv\chi=\dvv\psi^A\dvv\psi^B(\d_A\chi_B)$. Note that no further variables are in the kernel because by assumption the presymplectic structure is nondegenerate. Hence, we indeed reconstructed the initial  multisymplectic action~\eqref{multsymp2} and hence the symplectic structure is complete.

\subsection{Metric gravity}

A variety of standard examples of (gauge) field theories including e.g. (higher order) scalar field and Yang-Mills theory belong to the class of natural systems and their intrinsic Lagrangians were discussed in details already in~\cite{Grigoriev:2016wmk}.

Now we give another instructive example of Einstein gravity. Although the presymplectic formulation of gravity within presymplectic AKSZ framework was resently given in~\cite{Alkalaev:2013hta, Grigoriev:2016wmk,Grigoriev:2020xec} 
and is based on the Cartan-Weyl formulation in terms of the frame field and Lorentz connection it is worthwhile discussing the purely PDE theory framework employed in this work.

As a starting point we take usual Einstein-Hilbert action in the metric-like form:
\begin{equation}
S = \int d^\dmsn x \sqrt{-g}( R - 2 \Lambda)
\end{equation}
Lagrangian can be rewritten in the following form:
\begin{multline}
\label{laggrav_2}
\mathcal{L} = (dx)^n\sqrt{-g} (-\frac{1}{2}g^{\kappa \xi}{\d _\lambda}g_{\kappa \xi} g^{\alpha \beta} \Gamma ^\lambda _{\;\;\; \alpha \beta} - {\d _\lambda}g^{\alpha \beta} \Gamma ^\lambda _{\;\;\; \alpha \beta} + \frac{1}{2}g^{\kappa \xi}{\d _\alpha}g_{\kappa \xi} g^{\alpha \beta} \Gamma ^\lambda _{\;\;\; \lambda \beta} + \\
+ {\d _\alpha}g^{\alpha \beta} \Gamma ^\lambda _{\;\;\; \lambda \beta} + g^{\alpha \beta} \Gamma ^\gamma _{\;\;\; \alpha \beta} \Gamma ^\lambda _{\;\;\; \gamma \lambda} - g^{\alpha \beta}\Gamma ^\gamma _{\;\;\;\alpha \lambda} \Gamma ^\lambda _{\;\;\; \beta \gamma} - 2 \Lambda)\,,
\end{multline}
where 
\begin{equation}
    \Gamma^\gamma_{\alpha \beta} = g ^{\gamma \lambda} (\d _\alpha g _{\lambda \beta} + \d _\beta g_{\lambda \alpha} - \d _\lambda g _{\alpha \beta})\,,
\end{equation}
denotes the coefficients of the Levi-Civita connection determined by $g$. In this form the Lagrangian does not depend on the second derivatives of the metric tensor and it is easy to find the  presymplectic potential:
\begin{equation}
\label{chigrav}
\hat{\chi} = \sqrt{-g}(\Gamma ^{\rho \mu \nu} - \frac{1}{2}g^{\rho \mu}\Gamma _\lambda ^{\;\;\; \lambda \nu} - \frac{1}{2}g^{\rho \nu}\Gamma _\lambda ^{\;\;\; \lambda \mu} + \frac{1}{2}g^{\mu\nu}\Gamma _\lambda ^{\;\;\; \lambda \rho} - \frac{1}{2}g^{\mu \nu} \Gamma ^{\rho \lambda} _{\;\;\;\;\; \lambda})\dvv g_{\mu \nu} (dx)^{\dmsn}_\rho\,.
\end{equation}

The equation manifold $\cM$ (stationary surface) is determined by the Einstein equations and their total derivatives:
\begin{equation}
    R _{\alpha \beta} - \frac{1}{2} R g _{\alpha \beta} + \Lambda g _{\alpha \beta} = 0\,.
\end{equation}
Note that the equations do not constrain the metric and its first derivatives and hence $g_{\mu\nu}$, $\Gamma^{\lambda}{}_{\mu \nu}$ remain independent when restricted to the stationary surface. As coordinates on the equation  one can take $x^\mu$, $g _{\mu \nu}$, $\Gamma^{\lambda}{}_{\mu \nu}$ restricted to the stationary surface (by some abuse of notation we use the same notations for the coordinates restricted to the surface) along with those components of derivatives of $\Gamma^{\lambda}{}_{\mu \nu}$ that remain independent on $\cM$.  It this coordinate system the component expression of the pullback $\chi$ of the presymplectic potential $\hat\chi$ is given by exactly the same expression~\ref{chigrav}
while the covariant Hamiltonian reads as:
\begin{equation}
\mathcal{H} = \sqrt{-g}(\Gamma ^{\rho \mu \nu} \Gamma _{\nu\mu\rho} - \Gamma ^{\nu \mu} _{\;\;\;\;\;\mu} \Gamma ^\lambda _{\;\;\; \lambda \nu} + 2 \Lambda) (dx)^\dmsn\,.
\end{equation}
Finally, the intrinsic action can be written as:
\begin{multline}
S^C = \int d^\dmsn x \sqrt{-g} ({\d _\rho}g_{\mu \nu}(\Gamma ^{\rho \mu \nu} - \frac{1}{2}g^{\rho \mu}\Gamma _\lambda ^{\;\;\; \lambda \nu} - \frac{1}{2}g^{\rho \nu}\Gamma _\lambda ^{\;\;\; \lambda \mu} + \frac{1}{2}g^{\mu\nu}\Gamma _\lambda ^{\;\;\; \lambda \rho} - \frac{1}{2}g^{\mu \nu} \Gamma ^{\rho \lambda} _{\;\;\;\;\; \lambda}) - \\
- \Gamma ^{\rho \mu \nu} \Gamma _{\nu\mu\rho} + \Gamma ^{\nu \mu} _{\;\;\;\;\;\mu} \Gamma ^\lambda _{\;\;\; \lambda \nu} - 2 \Lambda)\,.
\end{multline}
In our coordinate system the intrinsic action only depends on $g _{\mu \nu}$ and $\Gamma ^{\lambda} {}_{\mu \nu}$. coordinates. Moreover, the presymplectic structure is nondegenerate in the sense of~\bref{sec:multisymp}. Thus, all other coordinates are to be gauged-away (for definiteness set to zero). By adding a total derivative (or, equivalently, picking a suitable presymplectic potential that determines the same presymplectic structure) it can be brought to the well-known Palatini form:
\begin{multline}
\label{iagrav}
S^C [g _{\mu \nu}, \Gamma ^\lambda {}_{\mu \nu}] = \int d^\dmsn x \sqrt{-g} g^{\mu \nu}({\d _\lambda}\Gamma ^\lambda _{\;\;\; \mu \nu} - \frac{1}{2}{\d _\mu}\Gamma ^\lambda _{\;\;\; \nu \lambda} - \frac{1}{2}{\d _\nu}\Gamma ^\lambda _{\;\;\; \mu \lambda} + \\
+\Gamma ^\gamma {} _{\mu \nu} \Gamma ^\lambda _{\;\;\; \gamma \lambda} - \Gamma ^\gamma _{\;\;\;\mu \lambda} \Gamma ^\lambda _{\;\;\; \nu \gamma} - 2 \Lambda)\,.
\end{multline}
Let us mention that an alternative presymplectic representation of the Einstein gravity action is based on the Cartan-Weyl formulation in terms of the frame field and Lorentz connection and has been proposed in~\cite{Alkalaev:2013hta} (see also~\cite{Grigoriev:2016wmk,Grigoriev:2020xec}).

To conclude the discussion of gravity in this formalism let us spell-out explicitly the gauge transformation:
\begin{equation}
\begin{gathered}
\delta g_{\alpha \beta} = \nabla _{(\alpha} \xi_{\beta)},\\
\delta \Gamma ^\lambda _{\;\;\; \mu \nu} = \frac{1}{2} g^{\rho \lambda} (\nabla _\mu \delta g_{\nu \rho} + \nabla _\nu \delta g_{\mu \rho} - \nabla _\rho \delta g_{\mu \nu})\,.
\end{gathered}
\end{equation}
These are just the standard transformations of metric tensor and its Levi-Civita connection under the infinitesimal diffeomorphisms.

\subsection{Fronsdal theory}

Consider as an additional example a theory of massless fields of arbitrary integer spin, known as Fronsdal theory. The Lagrangian for this theory reads as~\cite{Fronsdal:1978rb}:
\begin{multline}
\label{Fronsdal-L}
\mathcal{L} = (-\frac{1}{2}{\d ^\rho}\phi _{\mu(s)}{\d _\rho}\phi ^{\mu(s)} +\frac{1}{2} s {\d ^\nu} \phi _{\nu \mu(s-1)}{\d _\lambda} \phi ^{\lambda \mu(s-1)} + \frac{1}{4} s(s-1) {\d ^\rho}\phi ^\nu _{\nu \mu(s-2)}{\d ^\rho}\phi _\lambda ^{\lambda \mu(s-2)} - \\
 -\frac{1}{2}s(s-1){\d ^\rho}\phi _\nu ^{\nu \mu(s-2)} {\d ^\lambda} \phi _{\rho \lambda \mu(s-2)} + \frac{1}{8} s(s-1)(s-2) {\d ^\rho} \phi ^\nu _{\nu \rho  \mu(s-3)} {\d _\lambda} \phi _\tau ^{\tau \lambda \mu(s-3)})(dx)^\dmsn
\end{multline}
where $\phi_{\mu(s)}$ is a compact notation for the totally symmetric tensor field $\phi_{\mu_1\ldots \mu_s}(x)$ which is assumed double-tracelss.

It turns out that this system is also natural and the construction of the intrinsic action is completely standard. Leaving technical details to the Appendix~\bref{app:Fronsdal} we here only give an explicit expression for the intrinsic action for this theory:
\begin{multline}
\label{iaarbml}
S^C [\phi ^{\mu(s)}, \phi ^{\mu(s)} {}_{|\rho}] = \int d^\dmsn x (-{\d ^\rho}\phi ^{\mu(s)}\phi _{\mu(s)|\rho} + s {\d _\nu} \phi ^{\nu \mu(s-1)} \phi ^{\lambda} _{\mu(s-1)|\lambda} + \\
+  \ffrac{s(s-1)}{2} {\d _\rho}\phi ^\nu _{\nu \mu(s-2)} \phi _{\lambda} ^{\lambda \mu(s-2) |\rho} - \ffrac{s(s-1)}{2}{\d ^\lambda} \phi _{\rho \lambda \mu(s-2)} \phi _{\nu} ^{\nu \mu(s-2)|\rho} - \\
  - \ffrac{s(s-1)}{2} {\d _\rho}\phi ^\nu _{\nu \mu(s-2)} \phi _{\lambda} ^{\rho \mu(s-2)|\lambda } +\ffrac{s(s-1)(s-2)}{4} {\d ^\rho}\phi ^\nu _{\nu \rho \mu(s-3)} \phi _{\tau \lambda} ^{\tau \mu(s-3)|\lambda } - \cH ) \,,
\end{multline}
where
\begin{multline}
   \cH = (- \ffrac{1}{2}\phi  ^{\mu(s)|\rho}\phi _{\mu(s)|\rho} + \ffrac{s}{2} \phi _{\nu} ^{\mu(s-1) |\nu}\phi ^{\lambda} _{\mu(s-1) |\lambda} + \ffrac{s(s-1)}{4} \phi _{\nu} ^{\nu \mu(s-2)|\rho}\phi ^{\lambda} _{\lambda \mu(s-2)|\rho} - \\
  -\ffrac{s(s-1)}{2}\phi _{\nu} ^{\nu \mu(s-2) |\rho}\phi ^{\lambda} _{\rho\mu(s-2) |\lambda} + \ffrac{s(s-1)(s-2)}{8} \phi ^{\nu \rho} _{\nu \mu(s-3) |\rho}\phi _{\tau \lambda} ^{\tau \mu(s-3) |\lambda}) (dx)^\dmsn
\end{multline}
and an additional field $\phi^{\mu(s)|\rho}$ is assumed double-traceless in $\mu$ indexes. The gauge transformation for these fields can be written as:

\begin{equation}
\delta \phi _{\mu(s)} = \d _{(\mu} \lambda _{\mu(s-1) \;)}\,, \qquad \delta \phi _{\mu(s)|\rho} = \d _\rho \d_{(\mu} \lambda _{\mu(s-1) \;)}\,. 
\end{equation}

This action is a generalization of the linearized Palatini action to the case of higher spin fields. The formulation of Fronsdal theory in terms of presymplectic structures were considered in~\cite{Alkalaev:2013hta, Sharapov:2016qne, Sharapov:2021drr, Grigoriev:2012xg}. Let us  mention that it is different from an alternative first order action~\cite{Vasiliev:1980as}, known as frame-like action. That one can be naturally seen as a higher-spin generalization of the Cartan-Weyl action of gravity rather than Palatini action.

\subsection{Proca theory}

The equations of motion of massive spin $1$ field in Minkowski space read as:
\begin{equation}
\label{spin1-eoms}
\d _\mu \d ^\mu A^{\nu}+ m^2 A ^\nu = 0\,, \qquad {\d _\mu}A^\mu = 0\,.
\end{equation}
These equations can not directly arise as EL equations because the number of equations does not match the number of fields. However, it is well-known that there exists a Lagrangian that defines equivalent equations so that the above system should be regarded as a Lagrangian one. 

More specifically, the Lagrangian, is given by:
\begin{equation}
  \mathcal{L} = (-\frac{1}{4}F^{\mu\nu}F_{\mu\nu} + \frac{1}{2}m^2 A^\nu A_\nu)(dx)^\dmsn\,,
\end{equation}
and is known as Proca Lagrangian. Its EL equations read as
\begin{equation}
\d _\mu F^{\mu\nu}+ m^2 A ^\nu = 0\,.
\end{equation}
Applying $\d_\mu$ to both sides gives the second equation of~\eqref{spin1-eoms}. In other words the second equation of motion arises as a differential consequence of the EL equations.

It turns out that despite the differential consequences of lower order massive spin-1 theory is natural and the respective intrinsic Lagrangian is obtained in a straightforward way.
More precisely, action~\eqref{theory} defines a presymplectic potential:
\begin{equation}
\hat{\chi} = -\dvv A^\mu F^\nu _{\;\;\mu} (dx)^{n-1} _\nu
\end{equation}
As coordinates on the equation manifold it is convenient to take $x^\mu$, $A^\mu$, $F^\mu {}_\nu$, $S^{'\mu} {}_\nu$, $\ldots$ restricted to the surface and $S^{'\mu} _\nu$ denotes the symmetric traceless part of $D_\mu A^\mu _\nu$. In this coordinate system the expression for the $\hat\chi$ pulled-back to the surface reads as:
\begin{equation}
  \chi = -\dvv A^\mu F^\nu _{\;\;\mu} (dx)^{n-1} _\nu
\end{equation}
and the intrinsic action takes the form
\begin{equation}
\label{spin1-int}
  S^C[A,F] = \int d^\dmsn x (- \frac{1}{2}({\d ^\mu}A^\nu - {\d ^\nu}A^\mu) F_{\mu\nu} +\frac{1}{4}F^{\mu\nu}F_{\mu\nu} + \frac{1}{2}m^2 A^\nu A_\nu )\,.
\end{equation}
Here we assume that all the variables in the kernel of the symplectic structure have been already gauged-away.

It is easy to see that the presymplectic structure underlying~\eqref{spin1-int} is nondegenerate and $A,F$ remain independent on the stationary surface so that the system is indeed natural. To see that it is equivalent to the Proca action one considers EL equation for $F_{\tau\rho}$, giving $F^{\tau\rho} = {\d ^\tau}A^\rho - {\d ^\rho}A^\tau$. Eliminating $F^{\tau\rho}$
as an auxiliary field gives back the Proca action.


\section{Massive spin-2 field and its presymplectic structure}


We now turn to our central example of a system which is not a natural one. It is given by the spin 2 field in Minkowski space. The equations of motion can be written as follows:
\begin{equation}
    \label{em2m_4_1}
  ({\d _\lambda} {\d ^\lambda} - m^2 )\phi _{\mu \nu} = 0 
 \end{equation}
\begin{equation}
\label{em2m_4_2}
     {\d ^\mu}\phi _{\mu \nu} = 0\,,
\end{equation}
where $\phi _{\mu \nu}$ is assumed traceless and symmetric, i.e. $\phi _{\mu \nu} \eta^{\mu\nu}=0$ and $\phi_{\mu\nu}=\phi_{\nu\mu}$. 

The Lagrangian formulation for this system can not be constructed without introducing extra fields.  More precisely, 
in their celebrated work~\cite{Fierz:1939ix} Fierz and Pauli  added a new scalar field to the system and proposed a Lagrangian
whose equations of motion are equivalent to the above massive spin-2 equations of motion. More precisely, the extra scalar field is introduced by assuming $\phi_{\mu\nu}$ to be traceful (so that the initial field is identified as a trace-free component while the new one as the trace). The Lagrangian reads as 
\begin{multline}
\label{lag2m}
\mathcal{L} = (-\frac{1}{2}{\d _\lambda}\phi ^{\mu \nu}{\d ^\lambda}\phi _{\mu \nu} + {\d ^\mu}\phi _{\mu \nu}{\d _\lambda}\phi ^{\lambda \mu} + \frac{1}{2}{\d _\mu}\phi _\nu ^\nu{\d ^\mu}\phi _\lambda ^\lambda - {\d ^\lambda}\phi _{\lambda \mu}{\d ^\mu}\phi ^\nu _\nu - \\
-\frac{1}{2} m^2 (\phi ^{\mu \nu} \phi _{\mu \nu} - \phi ^\mu _\mu \phi ^\nu _\nu))(dx)^\dmsn\,,
\end{multline}
and is known as Fierz-Pauli Lagrangian.  We also assume that the space time dimension  $\dmsn$ is grater than $3$ and that $m \neq 0$. 

It is instructive to recall  how exactly the EL equations of the Fierz-Pauli Lagrangian reproduce the equation of motion and set to zero the additional scalar field. The crucial point is that the EL equations 
\begin{multline}
\label{em2m}
({\d _\gamma}{\d ^\gamma} - m^2) \phi _{\alpha \beta} - ({\d _\gamma}{\d ^\gamma} - m^2) \phi _\tau ^\tau \eta _{\alpha \beta} - \\
- {\d _\alpha} {\d ^\gamma} \phi _{\gamma \beta} - {\d _\beta} {\d ^\gamma} \phi _{\gamma \alpha} + {\d _\alpha} {\d _\beta} \phi _\gamma ^\gamma + {\d _\tau} {\d _\rho} \phi ^{\tau \rho} \eta _{\alpha \beta} = 0
\end{multline}
have  nontrivial differential consequences of lower order. More precisely applying $\eta^{\alpha \beta}$, ${\d ^\beta}$, and ${\d ^\alpha}{\d ^\beta}$ to \ref{em2m} gives respectively:
\begin{equation}
\label{em2m_1}
(2-\dmsn)({\d _\gamma}{\d ^\gamma} \phi _\alpha ^\alpha - {\d _\alpha} {\d _\beta} \phi ^{\alpha \beta}) + (\dmsn-1)m^2 \phi _\alpha ^\alpha = 0\,,
\end{equation}
\begin{equation}
\label{em2m_2}
{\d ^\beta} \phi _{\alpha \beta} - {\d _\alpha} \phi _\beta ^\beta = 0\,,
\end{equation}
and
\begin{equation}
\label{em2m_3}
{\d ^\alpha}{\d ^\beta} \phi _{\alpha \beta} - {\d ^\alpha}{\d _\alpha} \phi _\beta ^\beta = 0
\end{equation}

It is easy to see that  \eqref{em2m_1} and \eqref{em2m_3}
imply $\phi_\beta^\beta = 0$ which together with \eqref{em2m_2}) imply ${\d ^\beta} \phi _{\alpha \beta} = 0$. So that the addition field is indeed set to zero and the additional equation \eqref{em2m_4_2} is reproduced. Finally, taking into account $\phi_\beta^\beta = 0$ and ${\d ^\beta} \phi _{\alpha \beta} = 0$
the initial EL equation~\eqref{em2m} gives the remaining Klein-Gordon equation \eqref{em2m_4_1}. More formally, the prolongation of the EL equation of the Fierz-Pauli action coincides with the prolongation  of the spin-2 equation of motion and hence the spin-2 equations are to be regarded as the Lagrangian ones.

Now we attempt to construct an intrinsic Lagrangian starting from the Fierz-Pauli one. We use $\varphi_{\mu\nu}$ and $\varphi_{\mu\nu|\lambda}$ to denote respectively $\phi_{\mu\nu}$ and $\phi_{\mu\nu|\lambda}$ pulled back to the equation manifold $\cM$. Note that both $\varphi_{\mu\nu|\lambda}$ and $\varphi_{\mu\nu}$ are totally traceless thanks to 
traceless condition on $\varphi _{\mu \nu}$ and \eqref{em2m_4_2}. These coordinates can be completed to the following  coordinate system on $\cM$: $x^\mu$, $\varphi_{\mu \nu}$, $\varphi_{\mu\nu|\lambda}$, $\varphi _{{\mu \nu}|\lambda \rho} $, $\ldots$, where all the tensors can be assumed totally traceless thanks to the equations of motion.

The presymplectic potential determined by the Fierz-Pauli action can be chosen as
\begin{equation}
  \hat{\chi} = \dvv \phi ^{\mu \nu} (-\phi_{\mu \nu} {} ^{|\lambda}  + 2\phi_{\rho \nu} {} ^{|\rho} \eta ^\lambda _\mu + \eta _{\mu \nu} \phi ^{\rho |\lambda} _{\rho} - \frac{1}{2} \phi _{\rho |\nu} ^{\rho} \eta ^\lambda _\mu - \frac{1}{2} \phi _{\rho |\mu} ^{\rho} \eta ^\lambda _\nu  - \eta _{\mu \nu} \phi^{\rho \lambda} {} _{|\rho}) (dx)^{\dmsn-1} _\lambda\,.
\end{equation}
Using the coordinate system introduced above its pullback to $M$ reads as
\begin{equation}
  \chi = -\dvv \varphi ^{\mu \nu} \varphi_{\mu \nu}{}^{|\lambda}   (dx)^{\dmsn-1} _\lambda\,.
\end{equation}
The covariant Hamiltonian takes the form:
\begin{equation}
  \mathcal{H} = (-\frac{1}{2} \varphi^{\mu \nu} {} _{|\lambda}  \varphi_{\mu \nu} {} ^{|\lambda} + \frac{1}{2} m^2 \varphi ^{\mu \nu} \varphi _{\mu \nu})(dx)^\dmsn
\end{equation}
so that the intrinsic action reads as:
\begin{equation}
\label{ia2m}
 S^C = \int d^\dmsn x (-{\d _\gamma}\varphi ^{\alpha \beta} \varphi_{\alpha \beta} {} ^{|\gamma} + \frac{1}{2}\varphi^{\alpha \beta} {} _{|\gamma} \varphi_{\alpha \beta} {} ^{|\gamma} - \frac{1}{2} m^2 \varphi ^{\alpha \beta} \varphi _{\alpha \beta})
\end{equation}

Varying \ref{ia2m} with respect to $\varphi_{\mu \nu} {}^{|\lambda}$ and $\varphi_{\mu \nu}$ gives respectively
\begin{equation}
\label{add2m}
\varphi^{\alpha \beta} {} _{|\gamma} = {\d _\gamma}\varphi ^{\alpha \beta} - \frac{\dmsn}{\dmsn^2 + \dmsn - 2}({\d _\lambda}\varphi ^{\lambda \beta} \eta _\gamma ^\alpha + {\d _\lambda}\varphi ^{\lambda \alpha} \eta _\gamma ^\beta) + \frac{2}{\dmsn^2 + \dmsn - 2}{\d ^\lambda}\varphi _{\lambda \gamma} \eta ^{\alpha \beta}
\end{equation}
and 
\begin{equation}
\label{em2m_5}
{\d ^\gamma} \varphi^{\alpha \beta} {} _{|\gamma} - m^2 \varphi ^{\alpha \beta} = 0 \,.
\end{equation}
Substituting \eqref{add2m} into \eqref{em2m_5} we obtain
\begin{multline}
\label{em2m_6}
{\d ^\lambda} {\d _\lambda} \varphi ^{\mu \nu} - \frac{\dmsn}{\dmsn^2 + \dmsn - 2}({\d ^\mu}{\d _\lambda}\varphi ^{\lambda \nu} + {\d ^\nu}{\d _\lambda}\varphi ^{\lambda \mu}) + \frac{2}{\dmsn^2 + \dmsn - 2}{\d ^\gamma}{\d ^\lambda}\varphi _{\lambda \gamma} \eta ^{\mu \nu} - \\
- m^2 \varphi ^{\mu \nu} = 0
\end{multline}
These equations do not imply $\d_\mu \phi^{\mu\nu}=0$ so that the intrinsic Lagrangian does not give all the Fierz-Pauli conditions and hence the presymplectic structure is not complete. 
In other words, Fierz-Pauli equations give an example of a Lagrangian system whose Lagrangian is not encoded in the presymplectic structure on the equation manifold. It is tempting to conjecture that in this case the equation manifold is equipped with an additional geometric structure that together with the presymplectic one  determines the Lagrangian formulation. In the next section we attempt to give a certain characterization of this structure by explicitly constructing a minimal multisymplectic Lagrangian that is equivalent to the Fierz-Pauli one and can be considered as a proper extension of the above intrinsic Lagrangian.


\section{Minimal multisymplectic formulation}
\label{sec:parent}
\subsection{Parent action}

Given a Lagrangian system there is a systematic way to equivalently represent it in a multisymplectic form. 
This can be seen as a multidimensional version of the 
Ostrogradsky's action, see e.g. \cite{arxiv-1506.02210},
and is also known as a parent action.~\footnote{Despite the construction seems to be known we were not able to find an original  reference and hence refer to~\cite{Grigoriev:2010ic}
for details, further examples and generalizations.}  If, for simplicity, we restrict ourselves to Lagrangians of the form $\mathcal{L} = \mathcal{L}(\phi ^i, \phi ^i _{|\alpha}, \phi ^i _{|\alpha \beta})$ the parent action reads as:
\begin{equation}
\label{theory-Parent}
S^P[\phi,\phi_\alpha,\phi_{\alpha\beta},\pi _i ^{|\alpha},\pi _i ^{|\alpha \beta}] = \int d^\dmsn x (\mathcal{L} - \pi _i ^{|\alpha} ({\d _\alpha} \phi ^i - \phi  ^i _{|\alpha}) - \pi _i ^{|\alpha \beta} ({\d _\alpha}\phi ^i _{|\beta} - \phi ^i _{|\alpha \beta}))
\end{equation}
Introducing 
\begin{equation}
\overline{\chi} = \pi _i ^{|\alpha} d_F \phi ^i (dx) _\alpha ^{\dmsn-1}
\end{equation}
and
\begin{equation}
\overline{\mathcal{H}} = \dh \Psi ^i \overline{\chi} _i - \mathcal{L} (dx)^\dmsn = (\pi _i ^{|\alpha} \phi _{|\alpha} ^i - \mathcal{L}) (dx)^\dmsn\,,
\end{equation}
the parent action takes the manifestly multisymplectic form:
\begin{equation}
S^P = \int (\derham \Psi ^i \overline{\chi} _i - \overline{\mathcal{H}})\,.
\end{equation}

The the parent action is canonical because no auxiliary structures are employed in its construction. However, it is usually possible to eliminate some auxiliary field in such a way that the reduced  action (which is equivalent by construction) still has a multisymplectic form. Proceeding like this one arrives at the minimal multisymplectic formulation which can also be considered as a canonical one. At this stage the question is whether the undifferentiated field variables remain independent when restricted to the stationary surface. If they do, the system is natural and the multisymplectc action is equivalent to the intrinsic one.  If they do not, this means that the system is not natural and its EL equations have zeroth order differential consequences. These can be considered as
some sort of constraints in the multisymplectic formulation. Note however, that these constraints are not present in the case of $n=1$ (mechanics) and shoudl not be confused with constraints in the Hamiltonian formalism.

\subsection{Minimal action for massive spin 2}

Now we performed the above construction starting with the Fierz-Pauli action.  The parent action reads as:
\begin{multline}
\label{FP-parent}
S^P = \int d^\dmsn x (- \frac{1}{2} \phi _{\alpha \beta |\gamma} \phi ^{\alpha \beta |\gamma} + \phi _{\alpha \beta} {}^{|\alpha} \phi^{\gamma \beta} {}_{|\gamma}  + \frac{1}{2} \phi _{\alpha} ^{\alpha |\gamma} \phi _{\beta |\gamma} ^{\beta} - \phi^{\gamma \beta} {} _{|\gamma}  \phi _{\alpha |\beta} ^{\alpha} -
\\
- \frac{1}{2} m ^2 (\phi _{\alpha \beta} \phi ^{\alpha \beta} 
- \phi _\alpha ^\alpha \phi _\beta ^\beta) 
+\pi _{\alpha \beta |\gamma} ({\d ^\gamma} \phi ^{\alpha \beta} - \phi ^{\alpha \beta |\gamma}))\,.
\end{multline}

Now our goal is to eliminate a maximal number of the auxiliary fields whose elimination preserves the multysymplectic form of the action. For instance varying with respect to $\phi^{\mu \nu} {} _{|\lambda}$ gives us
\begin{equation}
\label{piphi}
\pi_{\mu \nu} {} ^{|\lambda} = - \phi_{\nu \mu} {} ^{|\lambda} + \phi_{\gamma \mu} {} ^{|\gamma} \eta _\nu ^\lambda + \phi_{\gamma \nu} {} ^{|\gamma} \eta _\mu ^\lambda + \phi _{\gamma} ^{\gamma |\lambda} \eta _{\mu \nu} - \phi^{\gamma \lambda} {} _{|\gamma} \eta _{\mu \nu} - \frac{1}{2} \phi _{\gamma |\mu} ^{\gamma} \eta _\nu ^\lambda - \frac{1}{2} \phi _{\gamma |\nu} ^{\gamma}\eta _\mu ^\lambda\,.
\end{equation}
This equation can be solved with respect to $\phi_{\mu\nu|\lambda}$ and hence $\phi_{\mu\nu|\lambda}$
is an auxiliary field. Indeed, taking traces of the above equation one gets
\begin{equation}
\pi _{\mu} ^{\mu |\lambda} = (\dmsn-2)(\phi _{\mu} ^{\mu |\lambda} - \phi^{\mu \lambda} {} _{|\mu})\,,
\qquad 
\pi^{\mu \lambda} {} _{|\mu}  = (1-\dmsn)(\frac{1}{2} \phi _{\mu} ^{\mu |\lambda} - \phi^{\mu \lambda} {} _{|\mu})
\end{equation}
This in turn leads to 
\begin{equation}
\label{pfi-pi-redef}
\phi_{\mu \nu} {} ^{|\lambda} = - \pi_{\nu \mu} {} ^{|\lambda} + \frac{1}{\dmsn - 1}\pi_{\gamma \mu} {} ^{|\gamma} \eta _\nu ^\lambda + \frac{1}{\dmsn-1}\pi_{\gamma \nu} {} ^{|\gamma} \eta _\mu ^\lambda + \frac{1}{\dmsn - 2}\pi _{\gamma} ^{\gamma |\lambda} \eta _{\mu \nu}\,.
\end{equation}

Substituting this back to the action~\eqref{FP-parent}
gives the minimal multisymplectic formulation of the system:
\begin{multline}
S = \int d^\dmsn x (- \frac{1}{2} \pi _{\alpha \beta |\gamma} \pi ^{\alpha \beta |\gamma} - \frac{1}{\dmsn-1} \pi_{\alpha \beta} {} ^{|\alpha} \pi^{\gamma \beta} {} _{|\gamma}  + \frac{1}{2(\dmsn-2) ^2} \pi _{\alpha} ^{\alpha |\gamma} \pi _{\beta |\gamma} ^{\beta} - \\
- \frac{1}{2} m ^2 (\phi _{\alpha \beta} \phi ^{\alpha \beta} - \phi _\alpha ^\alpha \phi _\beta ^\beta) + \pi _{\alpha \beta|\gamma} {\d ^\gamma} \phi ^{\alpha \beta} )\,.
\end{multline}
Note that if $\phi_{\mu\nu}$ and $\pi _{\alpha \beta |\gamma}$ were totally traceless this action would be identical  to the intrinsic one~\eqref{ia2m} provided one identifies $\pi _{\alpha \beta |\gamma}$ with $\phi _{\alpha \beta |\gamma}$ rescaled by a constant factor. 
However, the fields are traceful and the EL equations, as we expected and as we are going to demonstrate explicitly, have zeroth order differential consequences.

In this way we arrived at the explicit example of 
non-natural system and explicitly found its minimal multisymplectic form. Indeed,  the underlying symplectic structure is nondegenerate (in the sense of Section~\bref{sec:multisymp}) while, as we are going to see now, its equations of motion have nontrivial differential consequences of order zero so that $\phi^{\mu\nu},\pi^{\mu\nu}{}_{|\lambda}$
become dependent upon restricting to the equation manifold.

For the subsequent analysis it is convenient to perform an invertible field redefinition determined by
\eqref{piphi}. The resulting expression for the action take the form:
\begin{multline}
\label{almost_intrinsic}
S = \int d^\dmsn x ( - \phi ^{\mu \nu |\lambda} {\d _\lambda} \phi _{\mu \nu} + 2 \phi^{\lambda \mu} {} _{|\lambda}  {\d ^\nu} \phi _{\nu \mu} + \phi _{\mu} ^{\mu |\lambda} {\d _\lambda} \phi _\nu ^\nu - \phi^{\lambda \mu} {} _{|\lambda} {\d _\mu} \phi _\nu ^\nu - \phi _{\mu} ^{\mu |\lambda} {\d ^\nu} \phi _{\nu \lambda} 
+ 
\\
+\frac{1}{2} \phi _{\alpha \beta |\gamma} \phi ^{\alpha \beta |\gamma} - \phi_{\alpha \beta} {} ^{|\alpha} \phi^{\gamma \beta} {} _{|\gamma} - \frac{1}{2} \phi _{\alpha} ^{\alpha |\gamma} \phi _{\beta |\gamma} ^{\beta} + \phi^{\gamma \beta} {} _{|\gamma} \phi _{\alpha |\beta} ^{\alpha} -\frac{1}{2} m ^2 (\phi _{\alpha \beta} \phi ^{\alpha \beta} - \phi _\alpha ^\alpha \phi _\beta ^\beta))
\end{multline}
Note that the massless limit of this action is a special case of \eqref{iaarbml} with $s=2$ or a linearized case of \eqref{iagrav}.

Now we check that the EL equations are indeed equivalent to those of the Fierz-Pauli action and hence also have differential consequences of order zero. EL equations associated to $\phi_{\alpha \beta} {} ^{|\gamma}$ and $\phi_{\alpha \beta}$ read as
\begin{multline}
\label{add2ml}
-{\partial _\gamma} \phi ^{\alpha \beta} + {\partial _\lambda} \phi ^{\lambda \beta} \eta _\gamma ^\alpha + {\partial _\lambda} \phi ^{\lambda \alpha} \eta _\gamma ^\beta + {\partial _\gamma} \phi _\lambda ^\lambda \eta ^{\alpha \beta} - \frac{1}{2}({\partial ^\beta}\phi _\lambda ^\lambda \eta _\gamma ^\alpha + {\partial ^\alpha}\phi _\lambda ^\lambda \eta _\gamma ^\beta)  -\\
- {\partial ^\lambda}\phi _{\lambda \gamma} \eta ^{\alpha \beta} + \phi^{\alpha \beta} {} _{|\gamma} - \phi^{\lambda \beta} {} _{|\lambda} \eta _\gamma ^\alpha - \phi^{\lambda \alpha} {} _{|\lambda} \eta _\gamma ^\beta - \phi _{\lambda |\gamma} ^{\lambda} \eta ^{\alpha \beta} + \\
+\frac{1}{2}(\phi ^{\lambda |\beta} _{\lambda} \eta _\gamma ^\alpha + \phi ^{\lambda |\alpha} _{\lambda} \eta _\gamma ^\beta) + \phi_{\lambda \gamma} {} ^{|\lambda} \eta ^{\alpha \beta} = 0\,,
\end{multline}
\begin{multline}
\label{em2ml}
\d ^\gamma \phi^{\alpha \beta} {} _{|\gamma} - \d ^\alpha \phi^{\lambda \beta} {} _{|\lambda} - \d ^\beta \phi^{\lambda \alpha} {} _{|\lambda} - \d ^\gamma \phi _{\lambda |\gamma} ^{\lambda} \eta ^{\alpha \beta} + \d _\gamma \phi^{\lambda \gamma} {} _{|\lambda} \eta ^{\alpha \beta} + \frac{1}{2}(\d ^\alpha \phi ^{\lambda |\beta} _{\lambda} + \d ^\beta \phi ^{\lambda |\alpha} _{\lambda}) - \\
- m^2 \phi ^{\alpha \beta} + m^2 \phi _\gamma ^\gamma \eta ^{\alpha \beta} = 0
\end{multline}
Applying $\d^\gamma$ to the first equation and substituting it into the second equation one indeed arrives at EL equation \eqref{em2m} for the FP action and hence to the same  differential consequences $\d_\mu\phi^{\mu\nu}$ and $\phi^\mu{}_\mu=0$.

\subsection{Alternative representation of the action}

The relation between the intrinsic action (which is incomplete in our case) and the minimal multisymplectic one can be made more precise by reformulating  the latter in terms of the trace-free component fields and fields parameterizing the traces.  More precisely introducing  $\phi^{\lambda \mu} {} _{|\lambda} = \psi ^\mu$, $\phi _{\lambda |\mu} ^{\lambda} = \xi _\mu$, $\phi _\lambda ^\lambda = \rho$ and denoting tracefree components by $\varphi_{\alpha \beta}$ and $\varphi_{\alpha \beta|\gamma}$ the field redefinition takes the form:
\begin{multline}
\phi^{\alpha \beta} {} _{|\gamma} = \varphi^{\alpha \beta} {} _{|\gamma} + \frac{\dmsn}{\dmsn^2+\dmsn-2}(\psi ^\alpha \eta _\gamma ^\beta + \psi ^\beta \eta _\gamma ^\alpha) - \frac{2}{\dmsn^2+\dmsn-2}\psi _\gamma \eta ^{\alpha \beta} - \\
- \frac{1}{\dmsn^2+\dmsn-2}(\xi ^\alpha \eta _\gamma ^\beta + \xi ^\beta \eta _\gamma ^\alpha) + \frac{\dmsn+1}{\dmsn^2+\dmsn-2}\xi _\gamma \eta ^{\alpha \beta}
\end{multline}
The expression for the action in terms of the new variables read as
\begin{multline}
\label{prev6}
S = \int d^\dmsn x (- \varphi  ^{\alpha \beta|\gamma} {\d _\gamma} \varphi _{\alpha \beta}  + \frac{1}{2} \varphi   ^{\alpha \beta |\gamma} \varphi _{\alpha \beta|\gamma} -\frac{1}{2} m ^2 \varphi  _{\alpha \beta} \varphi  ^{\alpha \beta} + \\
+ \frac{2(\dmsn^2-2)}{\dmsn^2+\dmsn-2}\psi ^\beta {\d ^\gamma}\varphi _{\gamma \beta} - \frac{\dmsn^2+\dmsn-4}{\dmsn^2+\dmsn-2} \xi ^\beta {\d ^\gamma}\varphi _{\gamma \beta} + \frac{\dmsn-2}{\dmsn}\xi _\gamma {\d ^\gamma} \rho - \frac{\dmsn-2}{\dmsn}\psi _\gamma {\d ^\gamma} \rho - \\
- \frac{\dmsn^2-2}{\dmsn^2+\dmsn-2}\psi ^\alpha \psi _\alpha - \frac{\dmsn^2-3}{2(\dmsn^2+\dmsn-2)}\xi ^\alpha \xi _\alpha + \frac{\dmsn^2+\dmsn-4}{\dmsn^2+\dmsn-2}\psi ^\alpha \xi _\alpha + \frac{1}{2} m ^2 \frac{\dmsn-1}{\dmsn} \rho ^2)
\end{multline}
In Appendix~\bref{sec:A2} we explicitly relate this action to the Singh-Hagen form of massive spin-2 theory.

In the form~\eqref{prev6} it is obvious that the first three terms explicitly give the expression \ref{ia2m} for the intrinsic action.  The remaining variables can be thought of as coordinates on the fibers of the vector bundle of the stationary surface while the remaining terms can be interpreted in terms of certain geometric structures on the bundle. However, we postpone the investigation of this geometry to a future work.

\section{Massive spin-3 field}

Another example of a system which is not a natural one is the Lagrangian system of massive higher-spin fields~\cite{Singh:1974qz}.  In this work we limit ourselves to the case of spin-3 field as the Lagrangian formulation for all the higher spins is analogous but more involved technically. The Lagrangian for this theory reads as:\footnote{More precisely, this form is taken from~\cite{Rahman:2016tqc} and it is related to the original Singh-Hagen form~\cite{Singh:1974qz} through a field redefinition.}
\begin{multline}
\cL[\phi,\rho] = (-\frac{1}{2} \d _\gamma \phi ^{\mu \nu \lambda} \d^\gamma \phi _{\mu \nu \lambda} + \frac{3}{2} \d_\lambda \phi ^{\mu \nu \lambda} \d^\gamma \phi _{\mu \nu \gamma} + \frac{3}{4} \d_\gamma \phi ^{\mu \gamma} _\mu \d^\lambda \phi ^\nu _{\nu \lambda} +\frac{3}{2} \d_\nu \phi ^\gamma _{\gamma \mu} \d^\nu \phi ^{\lambda \mu} _\lambda - \\
- 3 \d^\mu \phi ^{\lambda \gamma} _\lambda \d^\nu \phi _{\mu \nu \gamma} - \frac{1}{2} m^2 \phi_{\mu \nu \gamma} \phi^{\mu \nu \gamma} + \frac{3}{2} m^2 \phi _{\nu \mu} ^\nu \phi ^{\lambda \mu} _\lambda +\frac{9}{4} m^2 \rho ^2 + \\
+ \frac{3(\dmsn - 1)(\dmsn - 2)}{2 \dmsn ^2} \d_\mu \rho \d^\mu \rho -\frac{3(\dmsn -2)}{2 \dmsn} m \rho \d_\mu \phi ^{\nu \mu} _\nu)(dx)^\dmsn\,,
\end{multline}
where $\phi_{\mu \nu \lambda}$ is the totally symmetric traceful field and $\rho$ is the scalar field. 

Let us recall how this Lagrangian reproduces the correct equations of motion. Just like in the of massive spin-2 considered above the EL equations for this Lagrangian
\begin{multline}
\label{em3m}
\Box \phi _{\mu \nu \lambda} - \d_\mu \d^\gamma \phi _{\gamma \nu \lambda} - \d_\nu \d^\gamma \phi _{\gamma \mu \lambda} - \d_\lambda \d^\gamma \phi _{\gamma \mu \nu} - \frac{1}{2} \eta _{\mu \nu} \d_\lambda \d^\gamma \phi _{\tau \gamma} ^\tau - \frac{1}{2} \eta _{\nu \lambda} \d_\mu \d^\gamma \phi _{\tau \gamma} ^\tau - \\
- \frac{1}{2} \eta _{\mu \lambda} \d_\nu \d^\gamma \phi _{\tau \gamma} ^\tau - \eta _{\mu \nu} \Box \phi _{\gamma \lambda} ^\gamma - \eta _{\nu \lambda} \Box \phi _{\gamma \mu} ^\gamma - \eta _{\mu \lambda} \Box \phi _{\gamma \nu} ^\gamma + \eta _{\mu \nu} \d^\gamma \d^\tau \phi _{\lambda \gamma \tau} + \eta _{\nu \lambda} \d^\gamma \d^\tau \phi _{\mu \gamma \tau} + \\
+ \eta _{\mu \lambda} \d^\gamma \d^\tau \phi _{\nu \gamma \tau} + \d_\mu \d_\nu \phi ^\gamma _{\gamma \lambda}  + \d_\lambda \d_\nu \phi ^\gamma _{\gamma \mu} + \d_\mu \d_\lambda \phi ^\gamma _{\gamma \nu} - m^2 \phi _{\mu \nu \lambda} + m^2 \eta_{\mu \nu} \phi ^\gamma _{\gamma \lambda} + \\
+ m^2 \eta_{\nu \lambda} \phi ^\gamma _{\gamma \mu} + m^2 \eta_{\mu \lambda} \phi ^\gamma _{\gamma \nu} + \frac{\dmsn - 2}{2\dmsn} m \eta_{\mu \nu} \d_\lambda \rho + \frac{\dmsn - 2}{2\dmsn} m \eta_{\nu \lambda} \d_\mu \rho + \\
+ \frac{\dmsn - 2}{2\dmsn} m \eta_{\mu \lambda} \d_\nu \rho = 0\,,
\end{multline}
\begin{equation}
\label{add3m}
\frac{3}{2} m^2 \rho - \frac{(\dmsn - 1)(\dmsn - 2)}{\dmsn^2} \Box \rho - \frac{\dmsn - 2}{2\dmsn} m \d_\mu \phi ^{\nu \mu} _\nu = 0\,.
\end{equation}
have nontrivial differential consequences. More precisely, the trace and the divergences of~\eqref{em3m} together with~\eqref{add3m} imply $\phi^\mu _{\mu \lambda} = 0$, $\rho = 0$ and hence
\begin{gather}
\label{em3m_1}
(\Box - m^2)\phi _{\mu \nu \lambda} = 0\,,\\
\label{em3m_2}
\d ^\mu \phi _{\mu \nu \lambda} = 0\,.
\end{gather}
Together with $\phi^\mu _{\mu \lambda} = 0$ these are precisely Fierz-Pauli conditions for the spin 3 massive field.

The equation manifold $\cM$ for the spin-3 system is determined by the prolongation of the \eqref{em3m_1} and \eqref{em3m_2}. As coordinates on $\cM$ it is convenient to take $x^\mu$, $\varphi _{\mu \nu \lambda}$, $\varphi _{\mu \nu \lambda|\gamma}$, $\ldots$, where all tensors are totally traceless. The computation of the presymplectic potential and the covariant Hamiltonian is straightforward and gives the following expression for the intrinsic action:
\begin{equation}
\label{ia3m}
 S^C = \int d^\dmsn x (-{\d _\gamma}\varphi ^{\mu \nu \lambda} \varphi_{\mu \nu \lambda} {} ^{|\gamma} + \frac{1}{2}\varphi^{\mu \nu \lambda} {} _{|\gamma} \varphi_{\mu \nu \lambda} {} ^{|\gamma} - \frac{1}{2} m^2 \varphi ^{\mu \nu \lambda} \varphi _{\mu \nu \lambda})
\end{equation}
It is easy to see that similarly to the spin-2 case the intrinsic action is not equivalent to the Singh-Hagen one.

We now construct a minimal extension of the intrinsic formulation, which has multisymplectic structure and is equivalent to the Singh-Hagen formulation. Following the same strategy as before we construct the parent formulation starting from the Singh-Hagen Lagrangian and then eliminate maximal amount of auxiliary fields without spoiling the multisymplectic form of the action. Here we only present the final result:
\begin{multline}
\label{almost_intrinsic_s3}
S = \int d^\dmsn x ( - \phi ^{\mu \nu \lambda} {} _{|\gamma} \d ^\gamma \phi _{\mu \nu \lambda} + 3 \phi ^{\lambda \mu \nu} {} _{|\lambda} \d ^\gamma \phi _{\mu \nu \gamma} + \frac{3}{2}\phi _\mu ^{\mu \gamma} {} _{|\gamma} \d _\lambda \phi _\nu ^{\nu \lambda} + 3 \phi ^\gamma _{\gamma \mu |\nu} \d^\nu \phi _\lambda ^{\lambda \mu} - \\
- 3 \phi _\lambda ^{\lambda \gamma |\mu} \d ^\nu \phi _{\mu \nu \gamma} - 3 \phi ^{\mu \nu \gamma} {} _{|\gamma} \d _\mu \phi ^\lambda _{\lambda \nu} + \frac{3(\dmsn - 1)(\dmsn - 2)}{\dmsn ^2} \rho ^{|\mu} \d _\mu \rho + \frac{3(\dmsn - 2)}{2\dmsn} m \phi ^\nu _{\nu \mu} \d ^\mu \rho +\\
+ \frac{1}{2} \phi _{\mu \nu \lambda} {} ^{|\gamma} \phi ^{\mu \nu \lambda} {} _{|\gamma} - \frac{3}{2} \phi ^{\lambda \mu \nu} {} _{|\lambda} \phi _{\gamma \mu \nu} {} ^{|\gamma} - \frac{3}{4} \phi _\mu ^{\mu \gamma} {} _{|\gamma} \phi ^\nu _{\nu \lambda} {} ^{|\lambda} - \frac{3}{2} \phi ^\gamma _{\gamma \mu |\nu} \phi _\lambda ^{\lambda \mu |\nu} + 3 \phi _\lambda ^{\lambda \gamma |\mu} \phi _{\gamma \mu \nu} {} ^{|\nu} - \\
- \frac{1}{2} m^2 \phi _{\mu \nu \lambda} \phi ^{\mu \nu \lambda} + \frac{3}{2} m^2 \phi _{\nu \mu} ^\nu \phi ^{\lambda \mu} _\lambda + \frac{9}{4} m^2 \rho ^2 - \frac{3(\dmsn - 1)(\dmsn - 2)}{2 \dmsn ^2} \rho ^{\mu} \rho _{\mu})\,.
\end{multline}
Details of the derivation are relegated to in Appendix~\bref{app:spin3}.
The above action depends on the traceful totally symmetric field $\phi_{\mu\nu\rho}$,  traceful $\phi^{\mu \nu \lambda} {} _{|\gamma}$ that is totally symmetric in first 3 indexes, vector file $\rho^{\mu}$, and scalar $\rho$.  This action is indeed an extension of~\eqref{ia3m} in the sense that if one requires $\phi_{\mu\nu\rho}$ and $\phi^{\mu \nu \lambda} {} _{|\gamma}$ to be totally traceless and $\rho$ and $\rho^{\mu}$ to vanish one arrives at~\eqref{ia3m}.

\section{Conclusion}
In this work we have studied presymplectic structures and intrinsic Lagrangians for massive fields in Minkowski space. We considered the first nontrivial case of massive spin 2 field and demonstrated that its natural intrinsic Lagrangian is not complete in the sense that it does not reproduce all the equations of motion. This feature is due to the presence of zeroth order differential consequences of the Euler-Lagrange equations of the known Fierz-Pauli Lagrangian or its natural first-order reformulation.  These consequences can be interpreted as an additional structure that together with the presymplectic structure defines an extended multisymplectic Lagrangian, equivalent to the Fierz-Pauli one. Such a Lagrangian is constructed as an equivalent reduction of the multidimensional version of the Ostrogradsky Lagrangian.

Despite that Lagrangians for massive fields can be easily constructed using St\"ueckelberg formalism or equivalently by dimensionally reducing massless Lagrangians in $n+1$ dimensions, these systems give nontrivial examples of theories where the natural symplectic structures do not encode all the equations of motion and hence are of utmost interest in the context of the Lagrangian formalism in field theory and the inverse problem of variations calculus. We showed that in a certain precise sense such systems are different from the natural ones for which the intrinsic Lagrangian can be made complete. 
In addition to the thorough discussion of massive fields and their multisymplectic formulations we reviewed the general construction of intrinsic Lagrangians in some details and illustrated it with various examples, including (constrained) mechanics and Fronsdal theory of massless higher spin fields.  We also consider Einstein gravity in the metric-like formulation and explicitly show that its intrinsic Lagrangian formulation is precisely the Palatini one.

Among possible further developments of this approach is its extension to a full scale BV-BRST formalism where the intrinsic Lagrangian formulation is promoted to a (generalized) presymplectic AKSZ sigma model. This construction has been recently put forward~\cite{Grigoriev:2020xec} in the case of gravity (see also~\cite{Alkalaev:2013hta,Grigoriev:2016bzl} for earlier relevant works and further examples). Closely related direction is to reanalyze the massive fields Lagrangians in the BV-BRST extension of the St\"ueckelberg formalism. Besides the development of the general formalism the results of this work should have potential applications to field theoretical models among which models of massive  gravity, see e.g.~\cite{Hinterbichler:2011tt,deRham:2014zqa} for a review, are of a particular interest.

\section*{Acknowledgments}
\label{sec:Aknowledgements}
The authors are grateful to I.~Khavkine and B.~Kruglikov for the illuminating discussion. M.G. also acknowledges useful discussions with M.~Henneaux, A.~Kotov, A.~Sharapov, and A.Verbovetsky. V.G. appreciates significant discussions with A.~Chekmenev and A.~Zimin. 
The work was supported in part by the Russian Science Foundation grant 18-72-10123. Part of this work was done when authors participated in the thematic program "Geometry for Higher Spin Gravity: Conformal Structures, PDEs, and Q-manifolds" at the Erwin Schrödinger International Institute for Mathematics and Physics, Vienna, Austria. Participation of V.G. was supported by Theoretical Physics and Mathematics Advancement Foundation “BASIS”.

\appendix
\section{Intrinsic action for Fronsdal theory: details of derivation}
\label{app:Fronsdal}
We start with the Lagrangian \eqref{Fronsdal-L}. Its EL equations read as
\begin{equation}
F_{\mu(s)} - \frac{1}{2} \eta _{(\mu \mu} F ^\nu _{\nu \mu(s-2) \;)}=0,
\end{equation}
where
\begin{equation}
F_{\mu(s)} = \Box \phi _{\mu(s)} - {\d _{(\mu}}{\d ^\nu} \phi _{\nu \mu(s-1) \;)} + {\d _{(\mu}}{\d _{\mu}} \phi ^\nu _{\nu \mu(s-2) \;)}
\end{equation}
As part of the coordinate on the stationary surface one can take 
$\phi _{\mu(s)}$ and $\phi _{\mu(s)|\rho}$ restricted to the surface. In this coordinates the presymplectic structure reads as

\begin{multline}
\chi = (- \phi_{\mu(s)} {} ^{|\rho} + s \phi _{\nu \mu(s-1)} {} ^{|\nu} \eta ^\rho _{\mu} + \frac{1}{2}s(s-1) \phi^\nu _{\nu \mu(s-2)} {} ^{|\rho} \eta _{\mu \mu} - \\
- \frac{1}{2}s(s-1) \phi^\nu _{\nu \mu(s-2)|\mu}  \eta ^\rho _{\mu}  - \frac{1}{2}s(s-1) \phi^\rho _{\lambda \mu(s-2)} {} ^{|\lambda} \eta _{\mu \mu} + \\
+ \frac{1}{4}s(s-2) (s-3) \phi^{\nu \lambda} _{\nu \mu(s-3)|\lambda} \eta_{\mu \mu} \eta ^\rho _{\mu})d_v \phi^{\mu(s)} (dx)^{\dmsn - 1} _\rho
\end{multline}
and the intrinsic actions takes the form~\eqref{iaarbml}

We now explicitly check that the intrinsic action is equivalent to the Fronsdal action, confirming that the Fronsdal theory is natural. The EL equations for $\phi^\rho {} _{\mu(s)}$ reads as:
\begin{multline}
-{\d ^\rho}\phi _{\mu(s)} + \eta ^\rho _{(\mu} {\d ^\nu} \phi _{\nu \mu(s-1) \;)} + \eta _{(\mu \mu}{\d ^\rho}\phi^\nu _{\nu \mu(s-2) \;)} - \eta _{(\mu \mu} {\d ^\lambda} \phi^\rho _{\lambda \mu(s-2) \;)} - \\
- \eta ^\rho _{(\mu} {\d _{\mu}} \phi^\nu _{\nu \mu(s-2) \;)} + \frac{1}{2} \eta ^\rho _{(\mu} \eta _{\mu \mu} {\d ^\lambda}\phi^\nu _{\nu \lambda \mu(s-3) \;)} + \phi_{\mu(s)} {} ^{|\rho} - \eta ^\rho _{(\mu} \phi_{\lambda \mu(s-1) \;)} {} ^{|\lambda} - \\
- \eta _{(\mu \mu} \phi^\lambda _{\lambda \mu(s-2) \;)} {} ^{|\rho}+ \eta _{(\mu \mu} \phi^\rho _{\lambda \mu(s-2) \;)} {} ^{|\lambda} + \eta ^\rho _{(\mu} \phi^{\nu} _{\nu \mu(s-2)|\mu \;)} - \\
- \frac{1}{2} \eta ^\rho _{(\mu} \eta _{\mu \mu} \phi ^\tau _{\tau \lambda \mu(s-3) \;)} {} ^{|\lambda}= 0
\end{multline}
By taking traces of these equations one finds ${\d ^\rho}\phi _{\mu(s)} = \phi_{\mu(s)} {} ^{|\rho}$. Substituting this back to the intrinsic action gives the initial Fronsdal action.

\section{Singh-Hagen form of the action}
\label{sec:A2}

We start with the multisymplectic formulation~\eqref{prev6}. Changing variables as $\psi^\mu \to \psi ^\mu - \frac{1}{\dmsn} \xi ^\mu$ the action takes the form
\begin{multline}
\label{prev7}
S = \int d^\dmsn x (- \varphi ^{\alpha \beta |\gamma} {\d _\gamma} \varphi _{\alpha \beta}  + \frac{1}{2} \varphi  ^{\alpha \beta|\gamma} \varphi _{\alpha \beta|\gamma}  -\frac{1}{2} m ^2 \varphi  _{\alpha \beta} \varphi  ^{\alpha \beta} + \\
+ 2\frac{\dmsn^2-2}{\dmsn^2+\dmsn-2}\psi ^\beta {\d ^\gamma}\varphi _{\gamma \beta} - \frac{\dmsn-2}{\dmsn} \xi ^\beta {\d ^\gamma}\varphi _{\gamma \beta} + \frac{(\dmsn-1)(\dmsn-2)}{\dmsn^2}\xi _\gamma {\d ^\gamma} \rho - \frac{\dmsn-2}{\dmsn}\psi _\gamma {\d ^\gamma} \rho - \\
- \frac{\dmsn^2-2}{\dmsn^2+\dmsn-2}\psi ^\alpha \psi _\alpha - \frac{1}{2}\frac{(\dmsn-1)(\dmsn-2)}{\dmsn^2}\xi ^\alpha \xi _\alpha + \frac{\dmsn-2}{\dmsn}\psi ^\alpha \xi _\alpha + \frac{1}{2} m ^2 \frac{\dmsn-1}{\dmsn} \rho ^2)
\end{multline}
Varying this action with respect to $\psi ^\alpha$ and $\xi ^\alpha$ we respectively obtain
\begin{equation}
-2\frac{\dmsn^2-2}{\dmsn^2+\dmsn-2} \psi _\alpha + \frac{\dmsn-2}{\dmsn} \xi _\alpha + 2\frac{\dmsn^2-2}{\dmsn^2+\dmsn-2} {\d ^\gamma} \varphi _{\gamma \alpha} - \frac{\dmsn-2}{\dmsn} {\d _\alpha} \rho = 0
\end{equation}
and 
\begin{equation}
-\frac{(\dmsn-1)(\dmsn-2)}{\dmsn^2} \xi _\alpha + \frac{\dmsn-2}{\dmsn} \psi _\alpha - \frac{\dmsn-2}{\dmsn} {\d ^\gamma} \varphi _{\gamma \alpha} + \frac{(\dmsn-1)(\dmsn-2)}{\dmsn^2} {\d _\alpha} \rho = 0\,.
\end{equation}
These imply $\psi _\alpha = {\d ^\gamma} \varphi _{\gamma \alpha}$ and $\xi _\alpha = {\d _\alpha} \rho$. Substituting this back into the action gives
\begin{multline}
\label{prev8}
S = \int d^\dmsn x (\frac{1}{2} \varphi  ^{\alpha \beta|\gamma} \varphi _{\alpha \beta|\gamma} -\frac{1}{2} m ^2 \varphi _{\alpha \beta} \varphi  ^{\alpha \beta} - \varphi ^{\alpha \beta |\gamma} {\d _\gamma} \varphi _{\alpha \beta} + \frac{\dmsn^2-2}{\dmsn^2+\dmsn-2}{\d _\alpha}\varphi ^{\alpha \beta} {\d ^\gamma}\varphi _{\gamma \beta} +\\
+ \frac{1}{2}\frac{(\dmsn-1)(\dmsn-2)}{\dmsn^2}{\d ^\gamma} \rho {\d _\gamma} \rho - \frac{\dmsn-2}{\dmsn} {\d ^\gamma}\varphi _{\gamma \beta} {\d ^\beta} \rho  + \frac{1}{2} m ^2 \frac{\dmsn-1}{\dmsn} \rho ^2)
\end{multline}
The EL equations for $\varphi^{\alpha \beta} {} _{|\gamma}$ read 
\begin{equation}
\label{add2m_1_n2}
\varphi^{\alpha \beta} _{|\gamma} = {\d _\gamma}\varphi ^{\alpha \beta} - \frac{\dmsn}{\dmsn^2 + \dmsn - 2}({\d _\lambda}\varphi ^{\lambda \beta} \eta _\gamma ^\alpha + {\d _\lambda}\varphi ^{\lambda \alpha} \eta _\gamma ^\beta) + \frac{2}{\dmsn^2 + \dmsn - 2}{\d ^\lambda}\varphi _{\lambda \gamma} \eta ^{\alpha \beta}
\end{equation}
Then substituting $\varphi^{\alpha \beta} {} _{|\gamma}$
back into \ref{prev8} gives the following action
\begin{multline}
S^{SH} = \int d^\dmsn x (- \frac{1}{2}{\d _\gamma}\varphi ^{\alpha \beta} {\d ^\gamma} \varphi _{\alpha \beta}  + {\d _\alpha}\varphi ^{\alpha \beta} {\d ^\gamma}\varphi _{\gamma \beta}  - \frac{\dmsn-2}{\dmsn} {\d ^\gamma}\varphi _{\gamma \beta} {\d ^\beta} \rho +\\
+ \frac{1}{2}\frac{(\dmsn-1)(\dmsn-2)}{\dmsn^2}{\d ^\gamma} \rho {\d _\gamma} \rho  -\frac{1}{2} m ^2 \varphi  _{\alpha \beta} \varphi  ^{\alpha \beta} + \frac{1}{2} m ^2 \frac{\dmsn-1}{\dmsn} \rho ^2)
\end{multline}
This is a Fierz-Pauli action in the Singh-Hagen form.\footnote{To obtain the standard form one should perform the following redefinition $\rho \rightarrow \frac{\dmsn}{\dmsn-2}\rho$.}

\section{Spin 3}
\label{app:spin3}

The Parent action \eqref{theory-Parent} for massive spin-3 theory can be written as follows

\begin{multline}
\label{spin3-Parent}
S^P = \int d^\dmsn x (- \frac{1}{2} \phi _{\mu \nu \lambda} {} ^{|\gamma} \phi ^{\mu \nu \lambda} {} _{|\gamma} + \frac{3}{2} \phi ^{\lambda \mu \nu} {} _{|\lambda} \phi _{\gamma \mu \nu} {} ^{|\gamma} + \frac{3}{4} \phi _\mu ^{\mu \gamma} {} _{|\gamma} \phi ^\nu _{\nu \lambda} {} ^{|\lambda} + \frac{3}{2} \phi ^\gamma _{\gamma \mu |\nu} \phi _\lambda ^{\lambda \mu |\nu} - \\
- 3 \phi _\lambda ^{\lambda \gamma |\mu} \phi _{\gamma \mu \nu} {} ^{|\nu} - \frac{1}{2} m^2 \phi _{\mu \nu \lambda} \phi ^{\mu \nu \lambda} + \frac{3}{2} m^2 \phi _{\nu \mu} ^\nu \phi ^{\lambda \mu} _\lambda + \frac{9}{4} m^2 \rho ^2 + \\
+ \frac{3(\dmsn - 1)(\dmsn - 2)}{2 \dmsn ^2} \rho ^{\mu} \rho _{\mu} + \frac{3(\dmsn - 2)}{2\dmsn} m \rho ^{\mu} \phi ^\nu _{\nu \mu} - \xi^{\mu} (\d _\mu \rho - \rho _{\mu}) - \\
- \pi _{\mu \nu \lambda} {} ^{|\gamma} (\d_\gamma \phi ^{\mu \nu \lambda} - \phi ^{\mu \nu \lambda} {} _{|\gamma}))
\end{multline}
The EL equations for  $\rho _{\mu}$ and $\phi _{\mu \nu \lambda} {} ^{|\gamma}$ read as
\begin{equation} \xi^{\mu} = - \frac{3(\dmsn - 1)(\dmsn - 2)}{\dmsn ^2} \rho ^{\mu} - \frac{3(\dmsn -2)}{2\dmsn} m \phi _\nu ^{\nu\mu}
\end{equation}
\begin{multline}\pi ^{\mu \nu \lambda} {} _{|\gamma} = \phi ^{\mu \nu \lambda} {} _{|\gamma} - \eta _\gamma ^\lambda \phi ^{\beta \mu \nu} {} _{|\beta} - \eta _\gamma ^\mu \phi ^{\beta \lambda \nu} {} _{|\beta} - \eta _\gamma ^\nu \phi ^{\beta \mu \lambda} {} _{|\beta} - \frac{1}{2} \eta ^{\mu \nu} \eta _\gamma ^\lambda \phi _\beta ^{\beta \alpha} {} _{|\alpha} - \frac{1}{2} \eta ^{\mu \lambda} \eta _\gamma ^\nu \phi _\beta ^{\beta \alpha} {} _{|\alpha} -\\- \frac{1}{2} \eta ^{\lambda \nu} \eta _\gamma ^\mu \phi _\beta ^{\beta \alpha} {} _{|\alpha} - \eta ^{\mu \nu} \phi _\beta ^{\beta \lambda} {} _{|\gamma} - \eta ^{\mu \lambda} \phi _\beta ^{\beta \nu} {} _{|\gamma} - \eta ^{\lambda \nu} \phi _\beta ^{\beta \mu} {} _{|\gamma} + \eta^{\mu \nu} \phi _\gamma ^{\lambda \beta} {} _{|\beta} + \eta^{\mu \lambda} \phi _\gamma ^{\nu \beta} {} _{|\beta} + \eta^{\lambda \nu} \phi _\gamma ^{\mu \beta} {} _{|\beta} + \\+ \frac{1}{2} \eta ^\mu _\gamma \phi _\beta ^{\beta \nu |\lambda} + \frac{1}{2} \eta ^\mu _\gamma \phi _\beta ^{\beta \lambda |\nu} + \frac{1}{2} \eta ^\nu _\gamma \phi _\beta ^{\beta \mu |\lambda} + \frac{1}{2} \eta ^\nu _\gamma \phi _\beta ^{\beta \lambda |\mu} + \frac{1}{2} \eta ^\lambda _\gamma \phi _\beta ^{\beta \nu |\mu} + \frac{1}{2} \eta ^\lambda _\gamma \phi _\beta ^{\beta \mu |\nu}
\end{multline}
These can be algebraically solved with respect to $\rho_\mu$ and $\phi _{\mu \nu \lambda} {} ^{|\gamma}$ and hence these variables are auxiliary and can be eliminated. At the same time these equations determine an invertible change of variables from $\xi^\mu$ to $\rho_\mu$ and from $\phi _{\mu \nu \lambda} {} ^{|\gamma}$ to $\pi _{\mu \nu \lambda} {} ^{|\gamma}$ so that one can equivalently eliminate $\xi^\mu$ and $\pi_{\mu \nu \lambda} {} ^{|\gamma}$, giving the action~\eqref{almost_intrinsic_s3}.

\bibliographystyle{utphys}
\bibliography{HSmaster}

\end{document}